\documentclass[twocolumn,tighten,times]{aastex62}

\newcommand{\Sec}[1]{Sect.~\ref{sec:#1}}
\newcommand{\Section}[1]{Section~\ref{sec:#1}}

\newcommand{\Fig}[1]{Fig.~\ref{fig:#1}}
\newcommand{\Figure}[1]{Figure~\ref{fig:#1}}

\newcommand{\Tab}[1]{Tab.~\ref{tab:#1}}
\newcommand{\Table}[1]{Table~\ref{tab:#1}}

\newcommand{\Eqn}[1]{Eqn.~(\ref{eq:#1})}
\newcommand{\Equation}[1]{Equation~(\ref{eq:#1})}

\newcommand\cm{\,\rm cm}

\newcommand\g{\,\rm g}

\newcommand\K{\,\rm K}

\newcommand\au{\,\rm AU}

\newcommand\tms{\!\times\!}
\newcommand\cdt{\!\cdot\!}
\newcommand\eq{\!=\!}

\newcommand\dt{\partial_t}

\newcommand\ee[1]{\tms 10^{#1}}

\newcommand\bb{\widehat{{\mathbf e}}_{\rm B}}

\newcommand\cs{c_{\rm s}}
\newcommand\vA{v_{\rm A}}
\newcommand\lH{l_{\rm H}}
\newcommand\qH{q_{\rm H}}
\newcommand\Rm{{\rm Rm}}
\newcommand\Rma{{\rm Rm_{art}}}
\newcommand\St{{\rm St}}

\newcommand\gmm{\gamma_{\rm m}}
\newcommand\kmax{k_{\rm m}}

\usepackage{mathtools}

\newcommand\V{\mathbf{v}}
\newcommand\B{\mathbf{B}}
\newcommand\J{\mathbf{J}}
\newcommand\Om{\mathbf{\Omega}}
\newcommand\omegaM{\boldsymbol{\omega}_{\rm M}}

\newcommand{\dB}[1]{\delta\mathbf {B}^{(#1)}}

\newcommand\vg{\mathbf{v}_{\rm g}}
\newcommand\vd{\mathbf{v}_{\rm d}}
\newcommand\vi{\mathbf{v}_{\rm i}}

\newcommand\rhog{\rho_{\rm g}}
\newcommand\rhod{\rho_{\rm d}}
\newcommand\rhodz{\rho_{{\rm d}\,0}}

\newcommand\Fdr{\mathbf{F}_{\rm drag}}

\newcommand\etaO{\eta_{\rm O}}
\newcommand\etaA{\eta_{\rm A}}
\newcommand\etaH{\eta_{\rm H}}
\newcommand\etaa{\eta_{\rm art}}
\newcommand\RO{r_{\rm O}}
\newcommand\RA{r_{\rm A}}
\newcommand\RH{r_{\rm H}}

\newcommand{\mn}[1]{\langle#1\rangle}

\newcommand{\simgt}%
           {\,\hbox{\lower0.35ex\hbox{$\sim$}\llap{\raise0.35ex\hbox{$>$}}}\,}
\newcommand{\simlt}%
           {\,\hbox{\lower0.35ex\hbox{$\sim$}\llap{\raise0.35ex\hbox{$<$}}}\,}

\usepackage{xspace}

\newcommand{\FTD}{\text{FARGO3D}\xspace}
\newcommand{\NIR}{\textsc{Nirvana-iii}\xspace}

\newcommand{\FTDs}{\textsc{F-3D}\xspace}
\newcommand{\NIRs}{\textsc{N-iii}\xspace}

\newcommand{\NBIA}{Niels Bohr International Academy, The Niels Bohr
  Institute, Blegdamsvej 17, DK-2100, Copenhagen \O, Denmark}

\newcommand{\KITP}{Kavli Institute for Theoretical Physics, University of California, Santa Barbara 93106, USA}

\newcommand{\DCU}{Centre for Astrophysics \& Relativity, School of Mathematical Sciences, Dublin City University (DCU), Ireland}

\usepackage{color,ulem,soul}

\definecolor{dark-red}{rgb}{0.75, 0.00, 0.00}
\definecolor{hlcolor}{rgb}{1.00, 0.90, 0.85}
\sethlcolor{hlcolor}

\received{May 31, 2018}\revised{July 30, 2018}\accepted{22 August, 2018}

\submitjournal{ApJ}

\graphicspath {}
\shorttitle{Hall-MHD simulations of protoplanetary disks}
\shortauthors{Krapp et al.}

\begin{document}

% ------------------------------------------------------------------------------

\title{\bf\Large
  Dust segregation in Hall-dominated turbulent protoplanetary disks}

\correspondingauthor{Leonardo Krapp}\email{krappleo@nbi.ku.dk}

\author{Leonardo Krapp}\affiliation{\NBIA}

\author{Oliver Gressel}\affiliation{\NBIA}\affiliation{\KITP}

\author{Pablo Ben\'itez-Llambay}\affiliation{\NBIA}

\author{Turlough P. Downes}\affiliation{\DCU}

\author{Gopakumar Mohandas}\affiliation{\NBIA}\affiliation{\KITP}

\author{Martin E. Pessah}\affiliation{\NBIA}

\begin{abstract}
  Imaging of the dust continuum emitted from disks around nearby protostars reveals diverse substructure.
  In recent years, theoretical efforts have been intensified to investigate how far the intrinsic dynamics of protoplanetary disks (PPDs) can lead to such features.
  Turbulence in the realm of non-ideal magnetohydrodynamics (MHD) is one candidate for explaining the generation of zonal flows which can lead to local dust enhancements.
  Adopting a radially varying cylindrical disk model, and considering combinations of vertical and azimuthal initial net flux, we perform 3D non-ideal MHD simulations aimed at studying self-organization induced by the Hall effect in turbulent PPDs.
  To this end, new modules have been incorporated into the \NIR and \FTD MHD codes.
  We moreover include dust grains, treated in the fluid approximation, in order to study their evolution subject to the emerging zonal flows.
  In the regime of a dominant Hall effect, we robustly obtain large-scale organized concentrations in the vertical magnetic field that remain stable for hundreds of orbits.
  For disks with vertical initial net flux alone, we confirm the presence of zonal flows and vortices that introduce regions of super-Keplerian gas flow.
  Including a moderately strong net-azimuthal magnetic flux can significantly alter the dynamics, partially preventing the self-organization of zonal flows.
  For plasma beta-parameters smaller than 50, large-scale, near-axisymmetric structures develop in the vertical magnetic flux.
  In all cases, we demonstrate that the emerging features are capable of accumulating dust grains for a range of Stokes numbers.
\end{abstract}
\vspace{-4ex}


% ------------------------------------------------------------------------------

\section{Introduction} \label{sec:intro}

Young stellar objects are found to harbor protoplanetary accretion disks (PPDs) composed of partially ionized gas  and dust grains inherited from the interstellar medium.
Renewed theoretical efforts have aimed at establishing a sound picture of the structure and evolution of these objects \citep[see][for a recent review]{Turner2014PPVI}.
These efforts are of paramount importance both for providing a framework in which theories of planet formation can be developed, and for aiding the interpretation of observations in the infrared and (sub-) millimeter wavebands \citep[e.g.,][]{2011ARA&A..49...67W}.

Observations appear to indicate that dust is settled-out to the disk midplane with a scale-height of $1\au$ at an orbital location of $100\au$ \citep{Pinte2016}.
In addition, they suggest that grains grow faster in the inner regions of the disk or, alternatively, that larger grains drift inward more efficiently.
In view of the growth of the embedded dust --- first to the mm-sizes detected in thermal emission in nearby PPDs and witnessed as chondrules in solar system meteorites, and then further to cm-sized pebbles and ultimately planetesimals --- it is important to consider the coupled aerodynamic evolution of the dust and the gas.

In featureless disks, intermediate-size dust grains are prone to rapid depletion due to their swift radial drift toward the star \citep{Whipple,Weidenschilling}.
Due to the drag force, gas pressure variations can slow down or stall the radial drift, creating regions of increased dust density, sometimes referred to as ``dust traps.''
Such pressure gradients can arise near the orbital location of (already formed) planets that are perturbing the gas disk, at opacity transitions, ice lines, dead-zone boundaries, or spontaneously via nonlinear feedback, as well as via self-organization of the flow into zonal flows or vortices \citep[see][for recent reviews]{Johansen2014, Testi2014PPVI}.

Outflows in the form of collimated protostellar jets and wide-angle disk winds \citep{Arce2007PPV,Bjerkeli2016} point toward the presence of dynamically relevant magnetic fields.
In particular, in the context of magnetized, turbulent disks resulting from the magnetorotational instability \citep[MRI;][]{Balbus1991}, pressure gradients can arise via the emergence of zonal flows \citep{Johansen2009}, or large-scale vortical flows \citep{Fromang2005,Johansen2005}.

Shearing box simulations of MRI in the ideal-MHD setting carried out by \citet{Johansen2009} showed that axisymmetric surface density fluctuations can grow and persist for many orbits.
Moreover, the resulting ``pressure bumps'' are typically found to be in geostrophic balance and are thus accompanied by  sub- / super-Keplerian rotation.
By means of 3D global simulations including Ohmic diffusion, \citet{Dzyurkevich2010} concluded that the excavation of gas from the active region during the linear growth and after the saturation of the MRI leads to the creation of a steady local radial gas pressure maximum near the dead zone edge, as well as to the formation of dense rings within the MRI-active region.

Beyond their very inner reaches (i.e., at radial distances from the star larger than a few tenths of an AU), PPDs are too cold for the gas to be thermally ionized \citep{DeschTurner2015}.
At the same time, from a few AU to a few tens of AU, the interior of the disk, near its midplane, is efficiently shielded from exterior sources of ionizing radiation \citep{TurnerDrake2009}, leaving the bulk of the gas in a state of partial ionization.
Thus, it is necessary to study the evolution of magnetic fields in the framework of \emph{non-ideal} MHD, where dissipative processes due to collisions are taken into account.

The central importance of the non-ideal effects has long been recognized  \citep[see, e.g.][]{Wardle1999,Sano2002,Salmeron2003}, but it is only recently that global simulations including all non-ideal effects --- i.e., Ohmic resistivity, ambipolar diffusion (AD) and the Hall term --- have been performed in relevant parameter regimes.
This has led to a modified picture of how PPDs accrete \citep{Bai2013, Gressel2015,Bai2017,Bethune2017}, a picture that deviates significantly from the traditionally envisioned dead zone \citep{Gammie1996}.

Detailed calculations of chemical abundances and ionization fractions using chemical reaction networks have provided an understanding of the relative importance of the three non-ideal effects \citep{Salmeron2008,Bai2011a}.
Ohmic resistivity becomes important in regions of high density and weak magnetic fields.
Those are typically expected to be located in the midplane region of the PPD, between approximately 1 and 5 AU.
The Hall effect is estimated to dominate close to the midplane, between 1 and 10 AU, in what is sometimes referred to as the planet-forming regions of PPDs.
AD, in turn, is anticipated to dominate in low-density regions with comparatively strong magnetic fields -- conditions applicable to the surface layers of the inner reaches of the disk, as well as the bulk of the outer disk (i.e., $\simgt 30\au$).

The role of AD in the formation of zonal flows and its potential impact on halting the inward drift of dust particles were first studied by \citet{Simon2014} using the local shearing-box approximation. Subsequently, using 3D unstratified global simulations, \citet{Zhu2014} found that turbulence is significantly suppressed around a planet-carved gap, and that a large vortex can form at the edge of the gap.
The vortex can efficiently trap dust particles that span three orders of magnitude in size within a timescale of about 100 orbits at the planet's location \citep{Zhu2015}.
Shearing box simulations including AD have been employed by \citet{Bai2014dust}, who found that magnetic flux can be concentrated into thin axisymmetric
shells, whose typical extent is less than half a pressure scale height.
Non-ideal global 3D MHD stratified simulations of the dead-zone outer edge \citep{Flock2015} have been analyzed in terms of potential observational signatures by \citet{Ruge2016}.
More recently, \citet{Suriano2017} have discussed the problem of dust trapping from the perspective of a magnetized wind in an AD-dominated disk.

In the presence of the Hall effect, the segregation of the turbulent flow into zonal bands appears to be even more pronounced.
For a local, vertically unstratified model, \citet{Kunz2013a} found that in Hall-dominated magnetorotational turbulence, zonal flows with radial concentrations of vertical magnetic flux develop.
The zonal flows are characterized by strong field amplitudes and are driven by a coherent Maxwell stress acting in concert with conservation of canonical vorticity --- a generalization of the flow vorticity motivated by the additional velocity appearing in the Hall-MHD induction equation.
Recent work by \citet{Bethune2016a} has confirmed the self-organization in a global cylindrical model, but still radially and vertically unstratified. The authors reported the generation of large-scale vortices and zonal flows, suggesting the possibility of dust trapping in the produced features.

Within the shearing-box framework, \citet{Lesur2014} showed that the Hall effect can induce a strong azimuthal field when vertical stratification is considered and zonal flows related to the local confinement of vertical magnetic flux can be inhibited.
\citet{Bai2015} also found zonal flows of vertical magnetic field in unstratified shearing box models; meanwhile stratified simulations show thin zonal flows that are supposedly not produced by the Hall effect.

The goal of our work is to assess the existence of dust traps in the form of zonal flows and vortices induced by Hall-MHD turbulence in protoplanetary disks.
We focus our current modeling efforts to the case of a vertically unstratified cylindrical disk considering only Ohmic dissipation and the Hall effect.
Such a simplification can be safely applied to the midplane of a typical PPD between $1 \textrm{ and } 5\au$, where the Hall effect likely dominates over Ohmic dissipation, and where AD is thought to be negligible.
We first aim at confirming the existing work and then extend the scope of the models to the dynamics of a relatively strong toroidal field ($B_{\phi0} \sim 10 B_{z0}$). We furthermore include dust fluids to rigorously confirm the ability of Hall-modified MRI turbulence to create self-organized features that are able to trap dust.

This paper is organized in the following manner. In \Section{model}, we introduce the equations of motion and the disk model, and in \Sec{methods} we describe our numerical methods. Results for a model without radial structure are presented in \Section{previous} and compared to existing work. In Sections \ref{sec:disk_self} and \ref{sec:dust}, we focus on the case of a radially varying disk ionization, where we moreover study the evolution of embedded dust grains. We conclude our discussion in \Section{conclusion}, and document our code implementation, as well as present different benchmarks in Appendices~\ref{sec:apA} and \ref{sec:apB}, respectively.

% ------------------------------------------------------------------------------

\section{Equations and disk model} \label{sec:model}

In the following, we describe the equations of motion as well as the underlying disk model that we employ.

\subsection{Equations} \label{sec:equations} % ---

We show below the continuity and momentum equations for the gas, and one of the dust species (we consider three dust species in total).
In addition we present the induction equation in a full non-ideal regime:
\begin{eqnarray}
\dt\rhog + \nabla\cdt (\rhog \vg) &\,=\,& 0\,, \nonumber\\[2pt]
\dt\rhod + \nabla\cdt (\rhod \vd) &\,=\,& 0\,, \nonumber\\[2pt]
\rhog (\dt\vg + \vg \cdt \nabla\vg) &\,=\,& -\!\nabla P - \rhog \nabla \Phi + \J \tms \B - \!\Fdr\,,\; \label{eq:mom_gas} \nonumber\\[2pt]
\rhod ( \dt\vd + \vd \cdt \nabla\vd ) &\,=\,&  - \rhod\nabla \Phi + \Fdr\,, \label{eq:mom_dust} \nonumber\\[2pt]
\dt\B &\,=\,& \nabla \times (\vg \tms \B) - \nabla \tms \mathcal{E}\,,
\end{eqnarray}
where $\rhog$ and $\vg$ are the density and velocity of the gas, whereas $\rhod$ and $\vd$ correspond to the dust density and velocity, respectively.
The term $\Fdr$ denotes the drag force between gas and dust, specified in \Equation{drag}. The non-ideal part of the electromotive force is given by
\begin{equation}\label{eq:EMF}
\mathcal{E} = \etaO \J \,+\, \etaH \J \tms \bb \,-\, \etaA \J \tms \bb \tms \bb\,,
\end{equation}
with $\bb$ the unit vector along $\B$, and where the current density is $\,\J \equiv \mu_0^{-1}\,\nabla \times \B$.

We furthermore adopt a locally isothermal equation of state where the disk temperature determines the sound speed $\cs$, given by $\cs^2(r) = P/\rhog$, with $P$ being the gas pressure.
The scalar variable $\Phi \equiv -GM_\odot/r$ is the cylindrical gravitational potential of a central solar mass star.

For the drag force acting on the dust, we consider the Epstein regime as introduced by \citet{Whipple} and \citet{Weidenschilling}, that is,
\begin{equation}  \label{eq:drag}
  \Fdr = \Omega_{\rm K}\,\St^{-1}\, ( \vg - \vd)\,,
\end{equation}
where $\Omega_{\rm K} = \sqrt{M_{\odot} G / r^3}$ is the Keplerian angular frequency and $\St \equiv \Omega_{\rm K}\,t_{\rm stop}$ is the so-called Stokes number.
In the context of the present work, $\St$ is considered to be constant, which implies a radial dependence of the stopping time, $t_{\rm stop}$.
The Epstein regime is valid as long as the mean free path of gas molecules is roughly larger than the particle radius. This is generally
true for grain sizes smaller than a few centimeters, when adopting the disk model of \Sec{disk}.
When including the drag force, the integration time of our simulations is short compared with the drift timescale.
Hence, we can safely neglect the back-reaction of the dust onto the gas.

Neglecting charges carried by grains, the Ohmic ($\etaO$), ambipolar ($\etaA$), and Hall ($\etaH$) diffusion coefficients can be expressed in terms of the species masses, $m$, number densities, $n$, and collision frequencies with neutrals, $\gamma$, as \citep{Bai2011a}
\begin{eqnarray}
  \etaO &\,=\,& \frac{c^2 \gamma_e m_e \rhog}{4 \pi e^2 n_e}
            \sim \frac{1}{x_e}\,, \nonumber \\[2pt]
  \etaH &\,=\,& \frac{c B}{4 \pi e n_e}
            \sim \frac{1}{x_e}\frac{B}{\rhog}\,, \nonumber \\[2pt]
  \etaA &\,=\,& \frac{ B^2}{4 \pi \gamma_i \rhog \rho_i}
  \sim \frac{1}{x_e} \bigg( \frac{B}{\rhog} \bigg)^2\,,
  \label{eq:scalings}
\end{eqnarray}
where $x_e = n_e/n$ is the ionization fraction.
It is common to quantify these terms using the Elsasser numbers
\begin{equation}
  \Lambda_{\rm O} \equiv \frac{v_A^2}{\etaO \Omega}\,,\qquad
  \Lambda_{\rm A} \equiv \frac{v_A^2}{\etaA \Omega}\,,\qquad
  \Lambda_{\rm H} \equiv \frac{v_A^2}{\etaH \Omega}\,,
\end{equation}
which characterize the coupling between the magnetic field and the neutral fluid at a length scale proportional to the characteristic wavelength of the MRI unstable modes \citep[e.g.][]{Wardle2012}.

Moreover, the relative importance of the Hall term can, in general, be defined via the so-called Hall length, $\lH  \equiv \etaH\, v_{\rm A}^{-1}$, \citep{Pandey2008,Kunz2013a}.
Adopting an incompressible non-stratified shearing box disk model, \citet{Kunz2013a} found that when $\lH/H\simgt0.2$ the MRI saturates to a new regime where the self-organization mechanism start to be operative.
The typical length $H$ can be adopted as the hydrostatic scale height of a vertical stratified disk.
Thus, $\lH/H$ provides a dimensionless control parameter for this new saturation regime of the MRI.

\subsection{Disk model}\label{sec:disk}

In \Section{disk_self}, we use the following disk model. The density and the sound speed are
\begin{equation}
  \rhog = \frac{\Sigma(r)}{\sqrt{2\pi}H(r)}\,,\quad
  \cs = c_{\rm s 0}\,r_{\text{AU}}^{-p} \;\,,
\end{equation}
with a surface density, $\Sigma(r)$, and the hydrostatic scale height, $H(r)$, defined as
\begin{equation}
  \Sigma(r) = \Sigma_0 r_{\text{AU}}^{-\sigma}
  \quad\text{and}\quad
  H(r) = \cs/\Omega_{\rm K}\,,
\end{equation}
respectively.
We initialize the vertical and azimuthal magnetic field components as
\begin{equation}
  B_{z0} = \displaystyle{ \sqrt{\frac{2 \mu_0 \rhog \cs^2}{\beta_z}}}\,\quad
  \text{and} \,\quad
  B_{\phi0} = \displaystyle{ \sqrt{\frac{2 \mu_0 \rhog \cs^2}{\beta_{\phi}}}}\,,
\end{equation}
respectively, where our units are such that $M_{\odot} = 1$, $\mu_0 = 1$. The radial distance, $r$, is given in \au. We typically quote elapsed time in terms of the orbital period, $t \equiv 2\pi \Omega^{-1}_{\rm K}(r_0)$ at the inner radius $r_0=1$ of the disk.

The initial plasma-$\beta$ parameter is set to be constant everywhere, $\Sigma_0 = 1.1\times10^{-4} M_{\odot}/r_0^2 \simeq 980\g\cm^{-2}$ and $c_{\rm s 0} = 0.05 \; r_0\Omega_{\rm K}(r_0)$.
We moreover choose $\sigma = 1$ as the power-law index for the surface density profile, and $p=1/2$ for the sound speed, yielding a constant aspect ratio of $h_0 \equiv H/r = 0.05$.

\subsection{Hall diffusion radial profile} % ---

For the purpose of complementing the basic cylindrical model with a radial structure, we assume a radially increasing ionization fraction.
In doing so, we stress the importance of maintaining consistency between the coefficients in terms of their dependence on $x_e$ as shown in \Eqn{scalings}.

In order to compare our model with previous results, we will make use of the Hall length, $\lH$, which in this simplified model is normalized by the initial density profile.
We assume
\begin{equation} \label{eq:etaH}
  \frac{\etaH}{|\B|} = q_{\rm H} \frac{h_0}{\sqrt{\rho_0}} \left(\frac{r}{r_0}\right)^{1+w} = \frac{\lH}{\sqrt{\rhog}}\,,
\end{equation}
where $\rho_0 = \Sigma_0 / \sqrt{2\pi}H(r_0)$ and $q_{\rm H} = \lH/H$ at $r = r_0$, that is,
\begin{equation} \label{eq:qH}
 q_{\rm H} = \frac{\lH(r_0)}{H(r_0)}.
\end{equation}
This defines a Hall length proportional to the aspect ratio of the disk, and a diffusion coefficient, $\etaH$, which is a function of the initial density profile.
In all our models, we use a vertical domain $L_z = 4H$, then $q_{\rm H}/4 = \lH(r_0)/L_z$.
With the ionization fraction an increasing power law of radius, that is, $x_e \sim r^u$, with $u>0$, we are looking for a limit on the $w$ exponent to establish a radial profile for the Hall diffusion.

Setting $w = 0.5$ and using equation~(10) from \citet{Kunz2013a}, we estimate an ionization fraction of
\begin{equation} \label{eq:gamma}
  x_e \simeq 3.9\times 10^{-12} \left(\frac{r}{r_0}\right)^{1/2}\,,
\end{equation}
yielding a maximum of $x_e \sim 8.7 \times 10^{-12}$ at radius $r = 5$.
We remark that we do not use a dedicated ionization model to compute $x_e$, but attempt to qualitatively describe radial profiles in the disk inner regions  (that is, up to two scale heights from the midplane).
The value obtained for the ionization fraction is in agreement with those presented by others \citep{Bai2009,Lesur2014,Keith2015,Bethune2017}, and it is at least one order of magnitude below the profile captured by \citet{RodgersLee2016}, implying that we are in a stronger Hall regime.

\section{Numerical methods} \label{sec:methods} % ---

We solve the non-ideal MHD equations in a cylindrical mesh with coordinates $(r, \phi, z)$.
We use extended versions of the {\FTD}\footnote{\href{http://fargo.in2p3.fr}{http://fargo.in2p3.fr}} \citep{Benitez-Llambay2016} and \NIR \citep{2004JCoPh.196..393Z,2011JCoPh.230.1035Z,2016A&A...586A..82Z} simulation codes --- the former including dust species in the form of pressureless fluids, and considering drag forces between the dust and gas (Ben\'itez-Llambay et al., in preparation).
We have also implemented new code modules embodying the Hall effect and AD.
In appendix~\ref{sec:apA}, we present a series of tests validating our implementations of the Hall scheme.
The majority of the results presented in this work have been obtained using \FTD, but we have taken the opportunity to inspect selected results with \NIR.
This is with the dual motivation of assessing potential numerical influences, as well as to cross-verifying the implementation of the Hall term in the two codes.

For solving the induction equation, both \FTD and \NIR use the constrained transport (CT) scheme, which guarantees $\dt \nabla\cdt\B\equiv0$ to machine precision \citep{Evans1988}.
To evolve numerically the magnetic field and compute the magnetic tension in the momentum equation, \FTD uses the method of characteristics (MOC) \citep{Stone1992}.
The Ohmic diffusion is integrated using a time-explicit update.

In both codes, we have implemented new modules in order to solve the Hall term following the approach presented by \citet{Bai2014} based on the operator-split solution of \citet{OSullivan2007}, termed Hall diffusion scheme (HDS). The \NIR code also implements the flux-based update suggested by \citet{2008JCoPh.227.6967T}.

The HDS scheme has been proven marginally stable under the Courant-Friedrichs-Lewy (CFL) condition $\Delta t < \Delta x^2/(2d\,\etaH)$, where $d$ is the dimension of the problem.
Our implementation generally appears robust in Cartesian and cylindrical coordinates.
Cell level noise is prone to appear if sharp gradients develop, but this is limited to the strong Hall regime, i.e., $\rm H/\rm I \equiv  |\etaH \J \times \bb| / | \vg \times \B| \sim 100$.

\subsection{Artificial resistivity} \label{sec:art} % ---

The Hall scheme implemented here is robust in terms of stability when the Hall effect does not strongly dominate the dynamics, i.e., $\rm H/\rm I\, \simlt 100$.
As pointed out before \citep[e.g.,][]{Huba2003,Bai2014}, in the case where the Hall term dominates, whistler waves can introduce spurious oscillations, affecting the stability of the numerical scheme.

We have attempted to remedy this issue by applying compact spectral filters \citep[\`a la][]{1992JCoPh.103...16L} on the edge-centered electric field within the CT step, but the approach has not been able to fully remove the grid-level noise.
In the context of solar physics, \citet{Vogler2005} have used a hyperdiffusivity scheme to improve the stability of their ideal-MHD scheme.
Adopting a similar strategy, we incorporate an artificial resistivity term that efficiently smooths strong gradients, removing cell-level noise without affecting the large-scale dynamics, as described below.

We note that the artificial resistivity is only required in the Hall-dominated regime.
Empirically, we have determined that it is sufficient to apply the artificial resistivity only near a maximum (or minimum) of $\B$ to obtain a stable evolution, resulting in a parameterization (at the grid location ${\bf x}_i$) of
\begin{equation}
  \etaa = {\eta^0_{\rm art}} \left( \frac{ \sum_s \text{max} (\dB{s}_x,\dB{s}_y,\dB{s}_z) }{ |\B|_{\rm ref} } \right)^{q_{\rm a}},
\end{equation}
where ${\eta^{0}_{\rm art}}$ has the dimension of diffusivity, and $s$ refers to the spatial direction (i.e., $x$, $y$ or $z$).
For instance, if we take $s=y$ and consider a cell with its center at the indexed position $(\phi_i,r_j,z_k)$, we have
\begin{eqnarray}
  \dB{y}_{x,i+1/2,j,k} = ( |B_{x,i+1/2,j,k} & - & B_{x,i+1/2,j-1,k}| \;+\; \nonumber \\[2pt]
  \;|B_{x,i+1/2,j+1,k} & - & B_{x,i+1/2,j,k}|\,)\;f(a)\,, \quad
 \label{eq:dBy}
\end{eqnarray}
with similar expressions for $\dB{x}_x$ and $\dB{z}_x$, respectively, but of course taking the difference along index $i$ and $k$.
Because we apply the diffusion only over a local maximum (minimum), we have
\begin{equation}
  f(a) \equiv \Bigg\{
  \begin{array}{lr}
    0 \quad\textrm{if}\quad a > 0 \\
    \frac{1}{2} \quad\textrm{if}\quad a \le 0
  \end{array}\Bigg. \,,
\end{equation}
with  $a\equiv (B_{x,i+1/2,j+1,k} - B_{x,i+1/2,j,k}) \times (B_{x,i+1/2,j,k} - B_{x,i+1/2,j-1,k})$, in the case of \Eqn{dBy}, and correspondingly for the other components.

We find that  ${\etaa}_0 \simeq 10^{-5}$ is typically sufficient to improve the stability of our numerical simulations.
A weighting of $q_{\rm a}=2$ helps to increase the contrast between regions where the artificial diffusion is needed or not, making the artificial dissipation more localized.
A seven-point stencil with $|\B|_{\rm ref} \equiv ( |\B|_{i,j,k} + |\B|_{i\pm1,j,k} + |\B|_{i,j\pm1,k} +|\B|_{i,j,k\pm1})/7$ proves to be convenient for the normalization.
The artificial resistivity term is also included when evaluating the CFL condition, i.e., we apply and effective Ohmic diffusion which is the sum of the physical and artificial terms, $\eta = \etaO + \etaa$.

The level of dissipation introduced by the artificial resistivity in our simulations can be measured via the magnetic Reynolds number $\Rm$, which
we define as
\begin{equation}
  \Rma \equiv \delta v_{\text{\rm rms}}\, L / \etaa\,,
\end{equation}
that is, taking the root mean square of the velocity perturbations, $\delta v_{\text{\rm rms}}$, as the characteristic speed.
We then compute the value of $\Rma$ setting $L$ as the vertical cell size $L = \Delta z \simlt r\Delta\phi < \Delta r$, where $r\Delta\phi$  and $\Delta r$ represent the azimuthal and radial cell sizes, respectively.
This choice for the length scale returns the minimum possible value for $\Rma$, providing a robust upper limit to the artificial dissipation.

In general, the artificial resistivity introduces values of $\Rma$ which are larger than $100$ for more than 80\% of the active domain, $\ge 10$ for 90\%, and never smaller than unity.
In all of our runs presented in \Sec{self}, we find that around 5\% of the active domain has a magnetic Reynolds number in the range $1<\Rma<5$.
This value drops below 3\%, when zonal flows develop.
In the case of the runs of Sections \ref{sec:disk_self} and \ref{sec:dust}, we include a background resistivity profile corresponding to a range of $10<\Rm<100$.
Introducing artificial resistivity in these models affects around 6\% of the active domain with $1<\Rma<5$.
We understand that such levels of dissipation have a minimal impact on the dynamics and do not affect the results that we are going to present.

\subsection{Sub-cycling the HDS step} \label{sec:subcyc} % ---

An extra complication is posed by the fact that, the permissible Hall-MHD time step might be significantly smaller than that for ideal MHD.
For that reason, we use a sub-cycling scheme that helps to speed up the integration.
Both in \FTD and \NIR, the sub-cycling integrates the magnetic field using the HDS scheme\footnote{In \NIR, when using the \citeauthor{2008JCoPh.227.6967T} scheme, no sub-cycling is used owing to the un-split character of the update.} with the Hall-MHD time step, producing an intermediate solution $B^*$.
Using $B^*$, we integrate the rest of the induction equation (advection + diffusion) with the time-step constraint from the CFL condition, but \emph{without} considering the Hall effect.
In the strong Hall regime, we find that the time step can be up to a hundred times smaller than the orbital advection time step.
Note that we include the FARGO / orbital advection scheme \citep{Masset2000,Stone2010} in both codes, reducing the computational integration time and the numerical diffusion.
As sub-cycling can potentially affect the propagation of high-frequency waves, we have performed additional convergence tests for oblique waves and the linear growth of the MRI (both local and global modes) with the sub-cycling enabled; all these tests are described in Appendix~\ref{sec:apB}.
For the oblique wave convergence test, we carried out runs with five and eleven sub-steps, respectively, and we generally obtained a slightly different error convergence rate, that is, up to 20\% smaller than that recovered without the sub-cycling.
In contrast, the linear growth of the MRI does not appear to be affected by the sub-cycling in our tests.
The HDS sub-cycling procedure typically decreases the computational time by a factor between three and five, depending on the adopted physical parameters.

\subsection{Super time stepping} % ---

When the sub-cycling scheme is applied for the HDS update, the time step is dominated by the Ohmic resistivity.
We then use a super-time-stepping (STS) scheme based on the solution of \citet{Alexiades} (both in \FTD and \NIR) and second-order Runge-Kutta-Legendre \citep[RKL2, ][]{2012MNRAS.422.2102M} in the case of \NIR.
We do not include a full test of the implementation here, but we have performed runs with and without the scheme for a few specific models, recovering nearly identical results.
In \NIR, we apply Strang splitting for the RKL2 step to maintain second-order accuracy in time. In \FTD, we update the magnetic field using the STS scheme after we apply the Hall scheme and before we solve the ideal-MHD induction equation, and the STS scheme follows the implementation described in \citet{Simon2013}.
The speed-up is within a factor of four in the case of \FTD for the more demanding simulations, where Ohmic diffusion induces a time step ten times smaller than the orbital advection step.

\subsection{Boundary conditions} \label{sec:boundaries} % ---

The numerical models presented here are all built to be periodic in the vertical and azimuthal directions.
We use reflecting radial boundaries for both the velocity and magnetic field.
The azimuthal velocity is extrapolated to the Keplerian profile, and we set the vertical velocity such that $\partial_r v_z = 0$.
Moreover, reflecting boundaries are also applied to the azimuthal ($\mathcal{E}_\phi$) and vertical ($\mathcal{E}_z$) components of the electromotive force (EMF), whereas for the radial component we have $\partial_r \mathcal{E}_r = 0$.
In combination, these boundary conditions conserve the magnetic and mass flux to machine precision while simultaneously preserving $\nabla\cdt\B = 0$.
On the downside, reflecting boundary conditions induce an accumulation of mass and vertical magnetic flux close to the radial boundaries when the MRI is fully developed.
To minimize the amount of  magnetic stresses at the radial boundaries, we define buffer zones (with a width of $0.2\,r_0$) where we apply Ohmic diffusion with a magnitude of $\etaO = 3 \times 10^{-4}$.
Inside this buffer zones we moreover define a region of size $0.05\,r_0$ where we restore the density to the initial density profile $\rho$.
In \FTD, following \citet{DeVal-Borro2006}, the density is modified as
$\rho \rightarrow (\rho\, \tau + \bar{\rho}_0 \Delta t) / (\Delta t + \tau)$, where $\tau \propto R(r) \Omega^{-1}$, with $R(r)$  a parabolic function that goes to one at the domain boundary and zero at the interior boundary of the wave-damping zone. In \NIR, we follow a similar procedure using the error function instead.

While the described procedure allows us to reduce the accumulation of mass at the inner boundary --- which, if left unchecked, would create undesired large-scale vortices --- it has the downside of generating a systematic drainage of the mass density.
In a typical simulation, we measure mass losses between 2\% -- 10\% of the initial mass contained in the domain.
The amount of mass lost increases when the saturation of the Hall MRI is reached. After that, the mass within the domain remains roughly constant, and we thus do not expect the overall findings to be severely affected.

% -------------------------------------------------------------------------------

\begin{table*}
  \centering
  \caption{Initial conditions and mesh parameters for the two models studied in \Sec{previous}.}
  \label{tab:compare}

  \begin{tabular}{lcccccccccc}
    \tablewidth{0pt}
    \hline\hline
    & $l_H $        & $\beta_{z}$            &  $\cs$               & $B_{z0}$        & ~ & $L_z\,(N_z)$ & $L_r\,(N_r)$ & $L_\phi\,(N_\phi)$
    &~& $\eta^{0}_{\rm art}$ \\
    \hline
    \decimals
    Weak Hall regime~~~ & $5.5\tms10^{-3}$  & $7.82\tms10^2$ &
                          $4.35\tms10^{-2}$ &  $2.2\tms10^{-3}$ &&
                          0.39 ~(36) & 4.2 ~(480) & $\pi/2$ ~(480) && --- \\
    Strong Hall regime  & $2.5\tms10^{-1}$  & $2.00\tms 10^4$ &
                          $1.00\tms10^{-1}$   &  $2.0\tms10^{-3}$ &&
                          0.25 ~(32) & 4.0 ~(512) & $\pi/2$ ~(256) && $5\ee{-5}$  \\
    \hline
  \end{tabular}
  \smallskip\newline
  {\footnotesize Domain sizes, $L_r$, $L_\phi$, and $L_z$ in the radial, azimuthal, and vertical direction, respectively, are listed with the number of cells given in brackets.}\medskip
\end{table*}

% -------------------------------------------------------------------------------

\subsection{Resolution requirements for MRI modes} \label{sec:resolution} % ---

We discuss below the resolution requirements for the linear growth of the MRI when combining Hall-MHD with Ohmic diffusion.
Despite the fact that our simulations are performed in a global cylindrical mesh, we make the simplified assumption that, for every radius, we can compute the local value of the maximum growth rate, $\gmm$, for the MRI.
We furthermore assume a disk with a Keplerian rotation profile, and
define $\kmax$ as the wavenumber with the maximum growth rate.

From linear theory, we obtain the following expression for the fastest growing wavenumber $\kmax$ \citep{Wardle2012,Mohandas2017}
\begin{equation}\label{eq:km}
        \kmax^2 = \frac{-4\gmm^2(\gmm^2 + \Omega^2)}
        {2v_A^2(2\gmm^2 \!-\! 3\Omega^2) \!-\! (\gmm^2 \!+\! \Omega^2)\left[3\Omega\etaH \!-\! 4\gmm\etaO\right]}\,,
\end{equation}
and the corresponding maximum growth rate, $\gmm$, in terms of the Hall diffusivity as
\begin{equation}\label{eq:etaHm}
  \etaH = \frac{24\,\Omega\,\etaO\gmm}{9\Omega^2 - 16\gmm^2}
  \,-\, \frac{2\Omega v_A^2}{\gmm^2 + \Omega^2}\,.
\end{equation}
If we only consider the Hall effect, equations (\ref{eq:km}) and (\ref{eq:etaHm}) can be combined to yield
\begin{equation}\label{eq:km2}
  \kmax^2 = \frac{2\,\Omega}{\etaH} \equiv \frac{2\sqrt{\!\beta}}{\lH H}.
\end{equation}
This result was obtained in the incompressible limit, but because of the low compressibility of our disk model, we can still establish the connection between the pressure scale height, $H$, and the wavenumber, $\kmax$, via \Eqn{km2}.

We guarantee a minimum resolution for the local linear instability that in general is around seven cells at $r_0$.
\citet{Hawley19952} suggested a minimum of five cells for codes that use finite difference schemes, such as \FTD and eight cells for shock capturing codes as \NIR.
While the effective resolution increases with radius, the maximum growth rate becomes smaller.
For instance, in the models F3D-bz1.4-bp0, and F3D-bz5.3-bp0 (see \Sec{disk}), the minimum resolution is eight cells at $r = r_0$ for the mode growing at a rate $\gamma_{\rm m} \simeq 0.68$.
The resolution is maximal around $r=4.5$, where $\lambda_{\rm m}/\Delta z \simeq 32$, and modes are reasonably resolved down to growth rates of $\gamma_{\rm m} \simlt 0.06$.

% ------------------------------------------------------------------------------

\section{Comparison with previous results} \label{sec:previous}

The aim of this section is to verify the implementation of our Hall-MHD schemes going beyond the linear tests presented in Appendix~\ref{sec:apB}.
More specifically, we focus on models that have minimal radial structure, i.e., where model parameters do not depend on the radial coordinate.
We begin our discussion by comparing global isothermal disk simulations with the existing results by \citet{Keeffe2014} and \citet{Bethune2016a}.
While the former was performed in the framework of multifluid MHD, the latter employed a single-fluid Hall-MHD approach ---
but both using a disk model where the vertical component of gravity was omitted.

In \Sec{updown}, we study the evolution of the dimensionless stress $\alpha$ (see Eq. \ref{eq:alpha}). This is done in regime where the Hall effect is comparable in magnitude to the ideal MHD term, constituting the so-called ``weak Hall regime''.
In this case, we do not make use of the artificial resistivity.

In \Sec{self}, we venture into a stronger Hall regime and we study the  self-organization process taking place within a Hall-MHD turbulent disk.
The initial conditions, domain size and number of cells are listed in \Table{compare} for reference.
The initial density is $\rho_0 = 1$.
In both models, we set a uniform initial magnetic field along the vertical direction and use a uniformly spaced cylindrical grid.

\subsection{Weak Hall regime: stress evolution} \label{sec:updown} % ---

\begin{figure}
  \centering\includegraphics[width=\columnwidth]{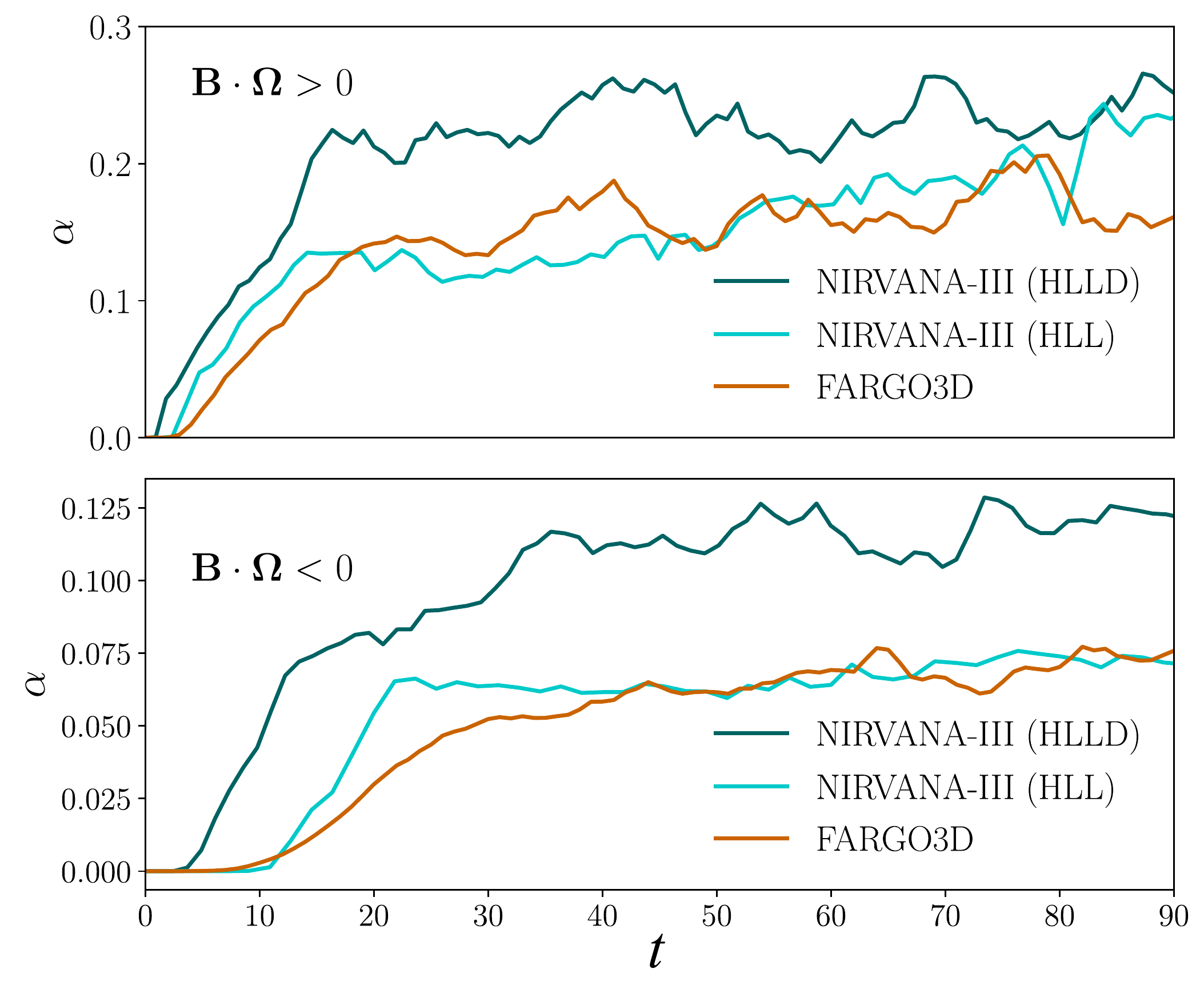}
  \caption{Time evolution of the dimensionless stress for the aligned (top panel) and anti-aligned configurations (bottom panel).}
  \label{fig:updown}
\end{figure}

It is well known that the dynamics of the Hall effect in a differentially rotating disk depend on the relative orientation between the angular velocity and the background vertical magnetic field \citep{Wardle2012, Kunz2013a, Bai2014}.
For a given range of the strength of the Hall effect, the resulting level of Maxwell stress can be amplified if both quantities are aligned, i.e., $\B\cdt\Om > 0$.
This behavior can be traced back to the linear regime, where the additional rotation of the MRI channel solution caused by the Hall current has a destabilizing effect compared to the ideal case \citep{Wardle2012}.

For the following discussion, we adopt the setup described in the Appendix~B of \citet{Bethune2016a}, which was developed with the aim of comparing the results with the previous work by \citet{Keeffe2014}.
The initial conditions are described in \Table{compare} (top row) and we run the setup using three different numerical schemes, that is,
\begin{enumerate}\itemsep0pt
  \item the unsplit HLL-based solver of \citeauthor{2008JCoPh.227.6967T},
  \item the split HLLD + HDS solver in \NIR,
  \item the MOC + HDS implementation in \FTD.
\end{enumerate}
For none of the runs, we made use of the artificial resistivity.
As can be checked from \Eqn{km2}, the fastest growing Hall-modified MRI mode, with $\lambda_{\rm m} \equiv 2\pi(H^2\lH^2/4\beta)^{1/4}$ is resolved with approximately three cells at $r=4$, which is not to the optimal resolution suggested in \Section{resolution}.

In \Figure{updown}, we show the evolution of the dimensionless stress, $\alpha$, with time, and we compute the time average, $\langle \alpha\rangle_{\rm T}$, of this quantity between $t=40$ and $t=80$. The dimensionless stress, $\alpha$, is defined as the sum of the $(r,\phi)$-component of the Reynolds and Maxwell stresses, normalized by the gas pressure.
We denote these two contributions with $\alpha_{\rm R}$ and $\alpha_{\rm M}$, respectively.
We normalize the stress as
$\alpha \equiv N_r^{-1} \sum_j \alpha_j$, where
\begin{eqnarray}\label{eq:alpha}
  \alpha_j & \,\equiv\, & \alpha_{\rm R} \; + \; \alpha_{\rm M}\,,
  \nonumber \\[4pt]
& = & \displaystyle{  \frac{ \sum_{i,k} (\rho v_r \delta v_\phi)_{i,j,k} }{c_{ {\rm s} j}^2 \sum_{i,k} \rho_{i,j,k} }
        - \frac{ \sum_{i,k} (B_r B_\phi)_{i,j,k}}{c_{{\rm s} j}^2 \sum_{i,k} \rho_{i,j,k} }}\,,
\end{eqnarray}
with indices $(i,j,k)$ as introduced in \Sec{art}, and where $\delta v_\phi$ is the deviation from the mean azimuthal velocity, and $N_r$ denotes the number of cells in the radial direction. Since the quantities in \Eqn{alpha} have, in general, different staggering, they need to be interpolated to the $r_j$ coordinate.

\begin{table}[t]
  \caption{Measurements of the mean $\alpha$-stress and magnetic energy}
  \label{tab:compare_results}

  \begin{tabular}{lcccc}
    \hline\hline
    Hall-MHD & $\langle \alpha\rangle_{\rm T+}$ & $\langle \alpha \rangle_{\rm T-}$ & $\langle E_B \rangle_{\rm T+}$ &  $\langle E_B \rangle_{\rm T-}$ \\[2pt]
    \hline
    \decimals
    \FTDs (HDS) & $0.17^{\pm0.02}$ & $0.066^{\pm0.004\!\!\!}$ & $0.32^{\pm0.03}$ & $0.12^{\pm0.01}$ \\
    \NIRs (HLL) & $0.15^{\pm0.02}$ & $0.067^{\pm0.005\!\!\!}$ & $0.22^{\pm0.02}$ & $0.12^{\pm0.01}$   \\
    \NIRs (HDS) & $0.23^{\pm0.02}$  &$0.12^{\pm0.01}$ & $0.30^{\pm0.04}$ & $0.17^{\pm0.02}$  \\[6pt]
    \hline\hline
    Ideal-MHD & $\langle \alpha\rangle $ & $\langle E_B \rangle$ &  \\[2pt]
    \hline
    \decimals
    \FTDs (HDS) & $0.11^{\pm0.01}$ & $0.19^{\pm0.02}$ \\
    \NIRs (HDS) & $0.16^{\pm0.01}$ & $0.25^{\pm0.03}$  \\
    \hline
  \end{tabular}
  \smallskip\newline
  {\footnotesize Averages between $t = 40-80$ for the different schemes.  The magnetic energy, $E_B$, is normalized to the initial pressure, i.e., $E_B \equiv B^2/(2\rho_0\, \cs^2)$.  Brackets $\langle.\rangle_{\rm T}$ indicate time averages and the sign $\rm T \pm$ refers to the aligned ($+$) and anti-aligned ($-$) configuration.}
\end{table}
The results are presented in \Table{compare_results}, where we can see that
the $\alpha$-values obtained with  \NIR using the HLL solver and \FTD only display a slight discrepancy in the average magnetic energy for the aligned configuration but otherwise agree well.

The $\langle \alpha \rangle_{\rm T}$ obtained with the MOC+HDS (for \FTD) and with HLL solver (for \NIR) are in agreement with those reported by \citet{Bethune2016a} but are larger than those reported by \citet{Keeffe2014}, where $\langle \alpha\rangle_{\rm T-} \sim 0.0092$ and $\langle \alpha\rangle_{\rm T+} \sim 0.075$ were obtained.

We moreover observe a difference between the time scales in which the MRI is found to saturate.
We obtain the saturation level after $t=20$ for $\langle \alpha\rangle_{\rm T+}$ and $t=30$ for $\langle \alpha\rangle_{\rm T-}$.
In contrast, \citet{Keeffe2014} obtained that the stress saturates between 10 and 15 inner orbits.
These differences might be related to the use of a different sound speed.
In addition, because of the single-fluid approach, our Hall diffusion coefficient only evolves with the magnetic field, whereas in the multifluid approach presented by \citet{Keeffe2014}, this coefficient is also updated with the densities and velocities of the charged species.
In view of this, the models are likely not directly comparable.

Regarding the magnetic energy evolution, we recover a similar trend as seen by \citet{Bethune2016a}.
The magnetic energy has the same saturation timescale as the $\alpha$-parameter, and remains stable around a constant mean value between the inner orbits 30 and 90.
This is another important difference with respect to \citet{Keeffe2014} where they found a secular growth as a function of time.

The $\langle \alpha \rangle_{\rm T}$ values obtained with \NIR using the HLLD solver are within 30\% larger compared with the ones from the other numerical schemes.
We conjecture that this is simply a consequence of the higher intrinsic accuracy of the HLLD scheme\footnote{See \citet{2010arXiv1003.0018B} for a detailed study on the role of Riemann solvers and slope limiters in the context of MRI in the ideal MHD limit.}, since
the discrepancy persists for measurements of $\alpha$ in ideal-MHD simulations. Using the same initial conditions as in \Table{compare_results}, but setting $\lH=0$, we found $\langle \alpha \rangle_{\rm T+} = 0.16$ with \NIRs, and  $\langle \alpha \rangle_{\rm T+} = 0.11$ for \FTDs.

After we have recovered the known dichotomy in the weak Hall regime with our Hall-MHD implementations, we now move on to a regime that involves a much stronger Hall effect.
Note that, in the following sections, we exclusively focus on the \emph{aligned} case, i.e., where $\B \cdt \Om >0$.
This is motivated by our interest in the strong Hall regime, that is, where $|\etaH| > 2\vA^2/\Omega$. When considering the \emph{anti-aligned} configuration in this region of parameter space, the Hall effect will always suppress the MRI \citep{Wardle2012} irrespective of the level of dissipation.
For this reason we exclusively focus on the \emph{aligned} case here. Note, however, that for weaker Hall effect, there exists a limited region of MRI growth in the anti-aligned configuration.

\subsection{Strong Hall regime: Self-organization\\
  of the vertical magnetic flux} \label{sec:self} % ---

\begin{figure}
  \begin{center}
    \includegraphics[width=\columnwidth]{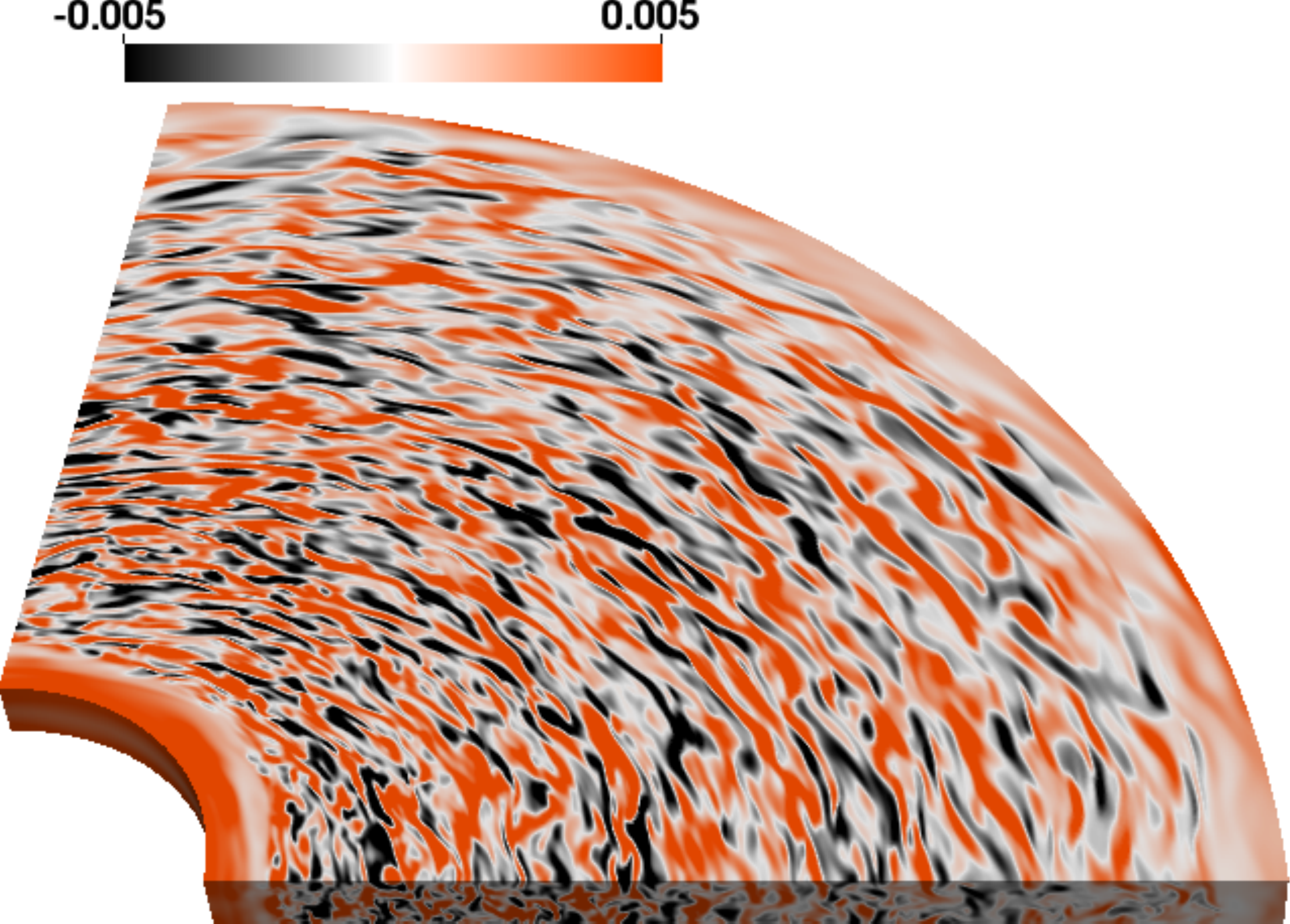}  \\[6pt]
    (a) $B_z$ after 50 orbits, before enabling the Hall effect.   \\[12pt]
    \includegraphics[width=\columnwidth]{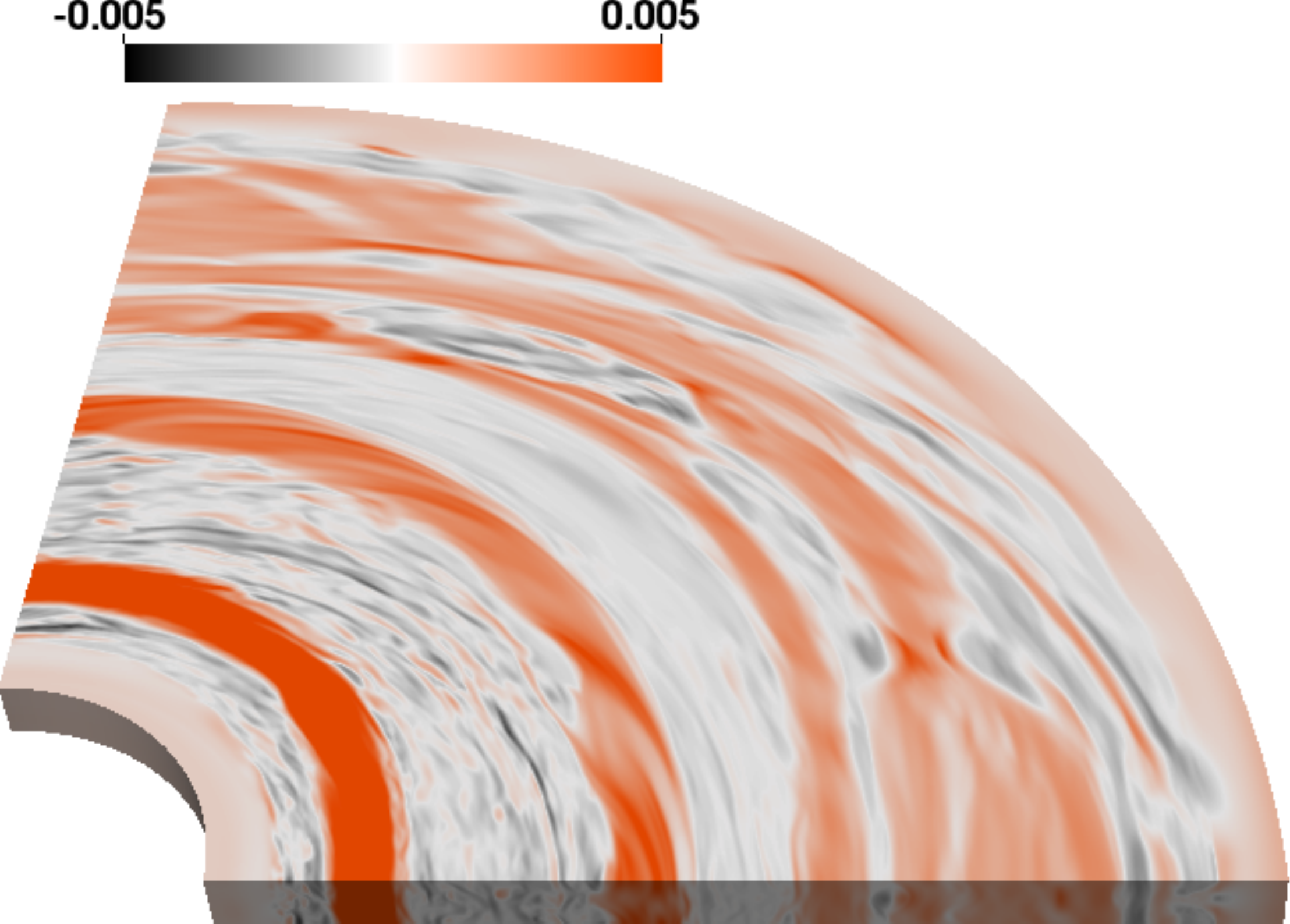} \\[6pt]
    (b) $B_z$ after 250 additional orbits with the Hall effect.                   \\
  \end{center}
  \caption{Evolution of Hall-MHD turbulence with $\lH = 0.25$. Color coding shows the vertical magnetic field.}
   \label{fig:self}
\end{figure}

\citet{Bethune2016a} performed an extensive study of self-organization by the Hall effect using a cylindrical disk model assuming globally constant initial conditions.
They conclude that a strong Hall effect is able to decrease the turbulent transport, in favor of producing large-scale azimuthal zonal flows accumulating vertical magnetic flux.
Depending on the plasma-$\beta$ parameter and the Hall length, these zonal flows can either evolve in the form of axisymmetric rings or vortices.
Their exploration of parameter space suggests that the general dynamics is not dramatically altered by the inclusion of Ohmic or ambipolar diffusion, nor by a non-zero toroidal magnetic flux.
\citet{Bethune2016a} moreover discuss the possibility that the vortices generated might eventually behave as dust traps via establishing regions of super-Keplerian rotation.
Their simulations, however, do not include dust directly, and
we will address this topic in \Section{disk} in the context of radially varying models.

In the following, we will begin by adopting the model B3L6 of \citet{Bethune2016a} as a starting reference (see the second line in \Table{runs}).
With these initial conditions, the vertical wavenumber $\lambda_{\rm m}$, yielding maximum linear growth for the Hall-MRI, is resolved with 22 cells at $r=2$  (also see \Sec{resolution}).
The detailed numerical procedure for the setup is described in \Sec{boundaries}, above.

In \Figure{self}, we show the vertical magnetic field at $t=50$, that is, \emph{before} turning on the Hall effect, and at $t=300$, when the fluctuations have undergone an extended phase of self-organization.
In our case, four bands of vertical magnetic flux are obtained, which is in good agreement with the model B3L6 described by \citet{Bethune2016a}.

\subsubsection{Confinement of magnetic flux} % ---

\begin{figure}
  \begin{center}
    \includegraphics[width=\columnwidth]{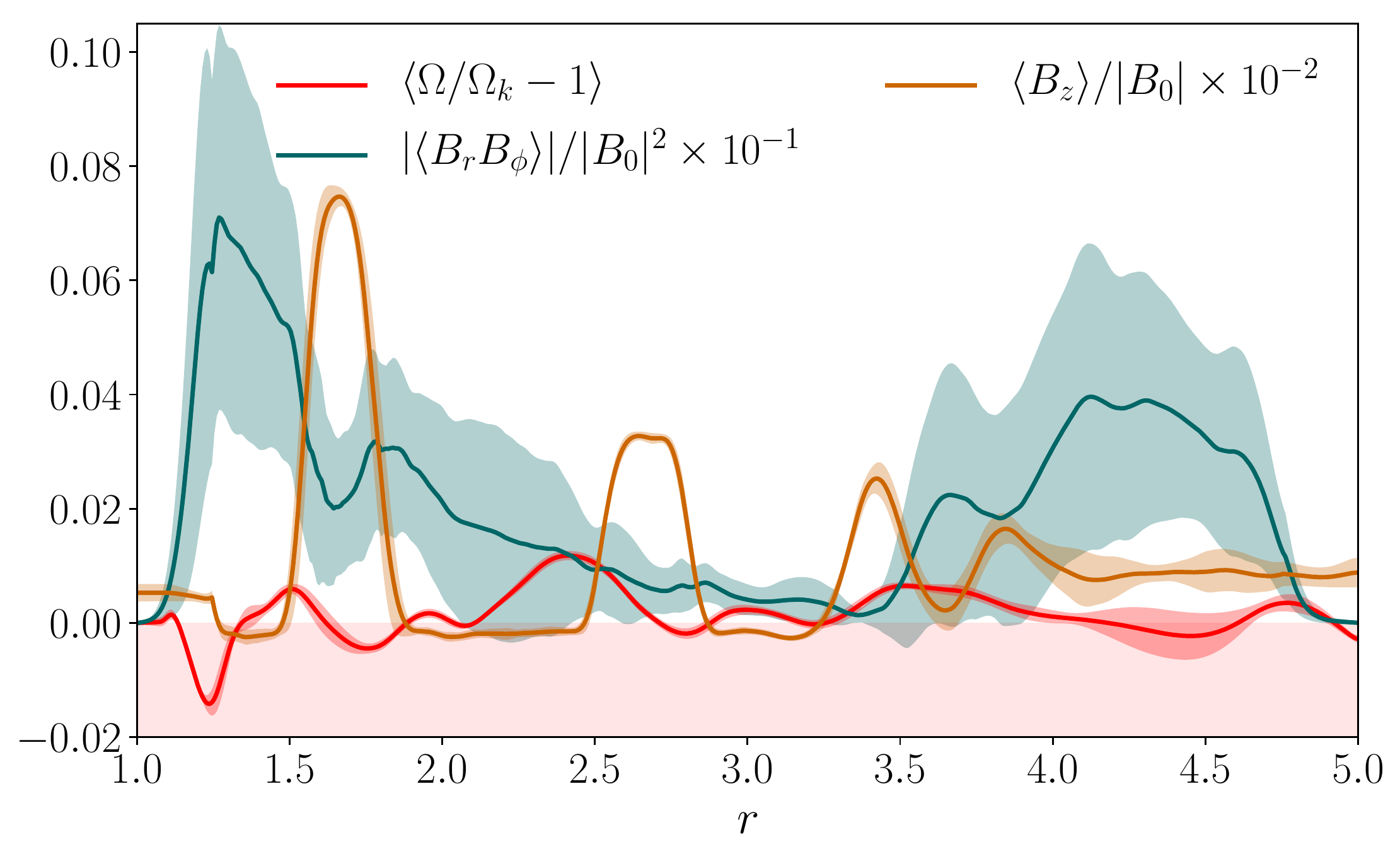} \\
   \end{center}
  \caption{Radial profiles of the vertical-azimuthal average of the vertical field, $\langle B_z \rangle $, the Maxwell stress, $\langle B_r B_\phi \rangle $, and $\Omega/\Omega_k - 1$. The solid curve correspond to the time average between $t=160 $ and $260$. Shaded areas indicate the standard deviation with respect to the time average. The light-red shaded area indicates the sub-Keplerian velocity region.} \label{fig:self_profile}
\end{figure}

In order to compare the stress level and the amount of flux confined in the zonal flows, we compute the radial profile $\langle B_z\rangle $ by taking vertical and azimuthal averages, normalized by the initial value $B_{z0}$.
We follow the same procedure for the absolute value of the Maxwell stress, $\langle B_r B_\phi \rangle $.
These quantities are plotted in \Fig{self_profile}, with the aim of showing how the vertical zonal flow is confined between regions of enhanced Maxwell stress. The values correspond to the time average between the $t=160 \,-\, 260$.

This mechanism of confinement was first explained by \citet{Kunz2013a} in the context of local shearing box simulations and was studied in a cylindrical domain by \citet{Bethune2016a}.
By means of a mean-field theory, they showed that the Hall effect introduces a component proportional to the turbulent Maxwell stress in the evolution equation of the vertical magnetic field.
This contribution of the Maxwell stress can be interpreted as an extra dissipation (of either sign) added to the Ohmic diffusion --- in other words, it can favor the accumulation of magnetic flux in regions of positive curvature of the mean Maxwell stress.
Once the zonal flow is formed, the turbulent field in-between starts to flow into the regions of confinement.
This generates a region of low stress and vanishing vertical flux between the azimuthal bands.
On the other hand, turbulent fluctuations may survive in regions located close to the zonal flows, which can also be seen in \Fig{self_profile}.

After around 300 orbits, the zonal flows remain quasi stationary, and
three separate zonal flows are clearly distinguishable, with a fourth one seeming to appear close the outer boundary at $r\sim4$.
\Figure{self_profile} also shows that, on average, the Maxwell stress is larger near the radial buffer zones.
Reflecting boundary conditions adopted in the radial direction (enabling conservation of the magnetic flux to machine precision) bring about the accumulation of magnetic flux close to the buffer zones, and this may eventually favor the concentration of Maxwell stress as well.

\subsubsection{Effect on the azimuthal velocity}

In addition to the averages of the Maxwell stress and the vertical field, we plot in \Figure{self_profile} the vertically, azimuthally and time averaged radial profile of $\Omega/\Omega_K - 1$ indicating deviations of the velocity field from Keplerian rotation (the light-red shaded region indicates sub-Keplerian rotation).
The emerging zonal flows may introduce regions of super-Keplerian rotational velocity which, in turn, may act locally to accumulate dust grains.
We find in our run that the velocity can reach a robust super-Keplerian regime, constituting an efficient dust trap, that will prospectively produce an azimuthally large-scale dust distribution.
Note, however, that in the absence of a radial pressure gradient in the background flow --- as assumed in the models presented in this section --- the gas initially rotates at the Keplerian speed.
In contrast, in a real disk, stellar irradiation heating will lead to a radial pressure profile, which will, in general, decrease with radius.
In that situation, the gas disk will rotate on a slightly sub-Keplerian rotation profile, requiring a more vigorous effect to achieve regions of super-Keplerian rotation.

% ------------------------------------------------------------------------------

\section{Self-organization in models\\ with radial disk structure} \label{sec:disk_self} % ---

\begin{figure}
  \centering\includegraphics[width=1.\columnwidth]{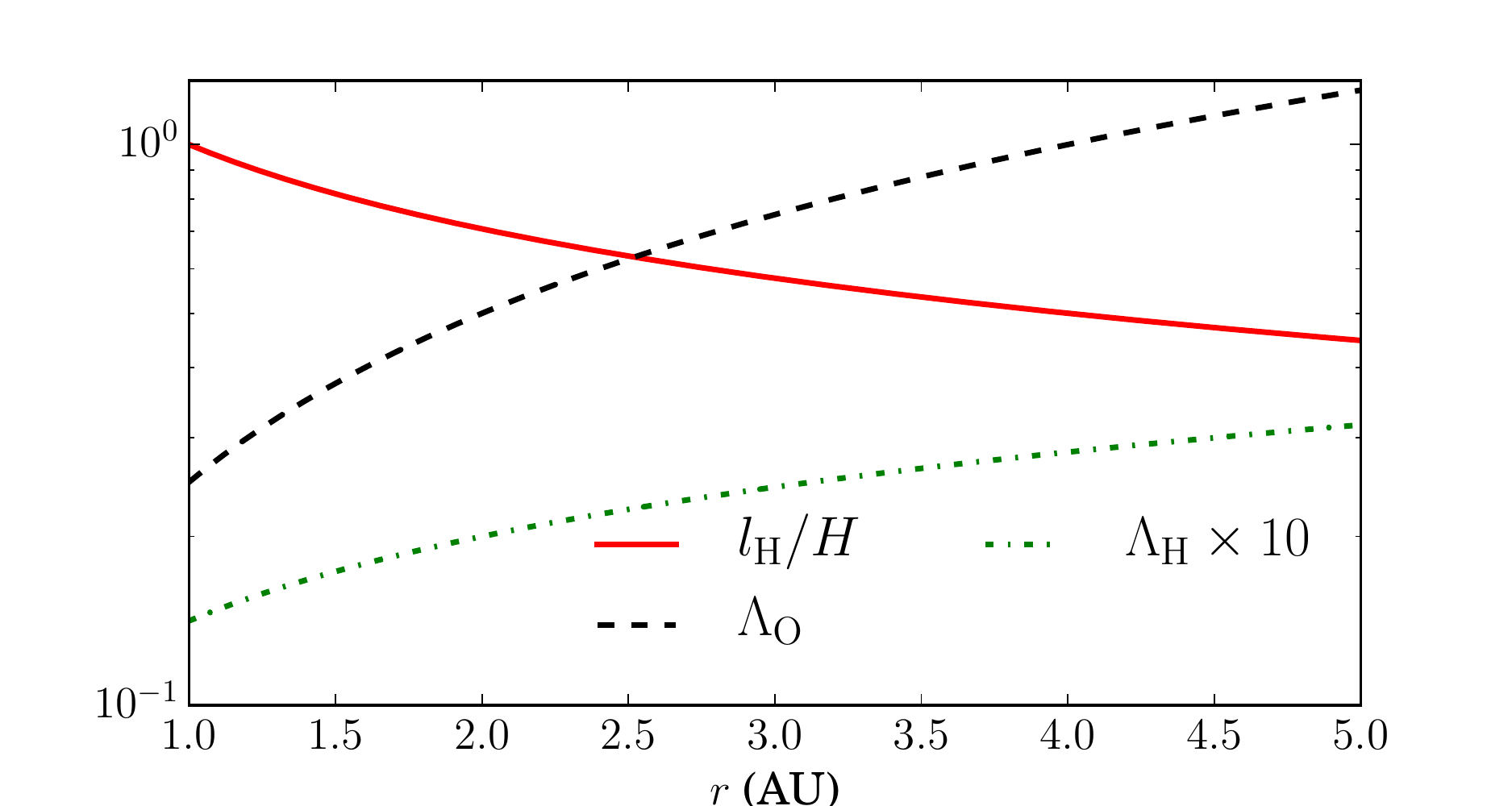}
  \caption{Elsasser numbers, $\Lambda_{\rm O}$, and $\Lambda_{\rm H}$, as well as $\lH/H$ for $\beta_z = 10^{4}$ and $\beta_{\phi}=0$, corresponding to model F3D-bz1.4-bp0} \label{fig:elsasser}
\end{figure}

\begin{figure}
  \includegraphics[width=0.83\columnwidth]{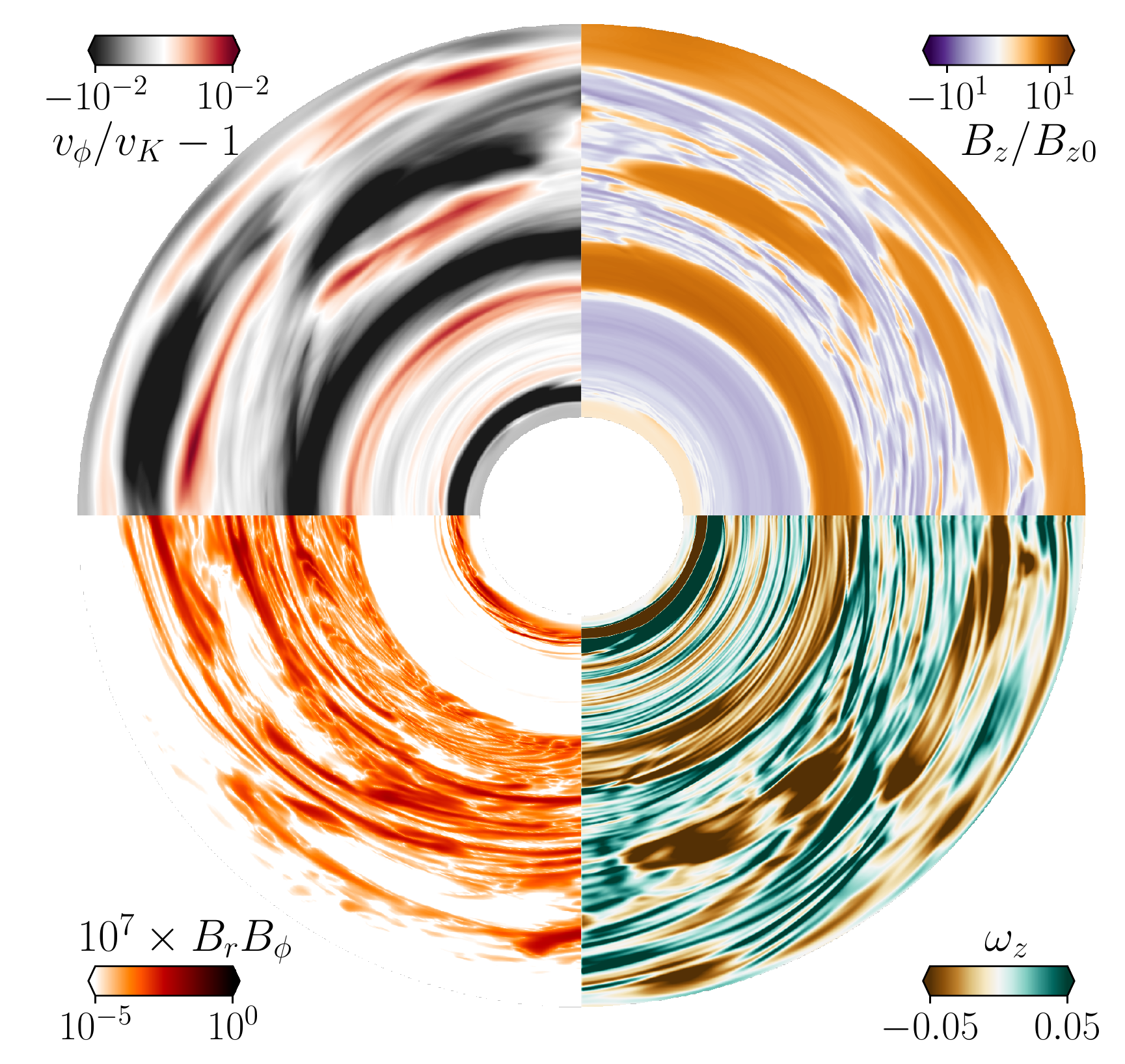}  \\[3pt] \includegraphics[width=0.83\columnwidth]{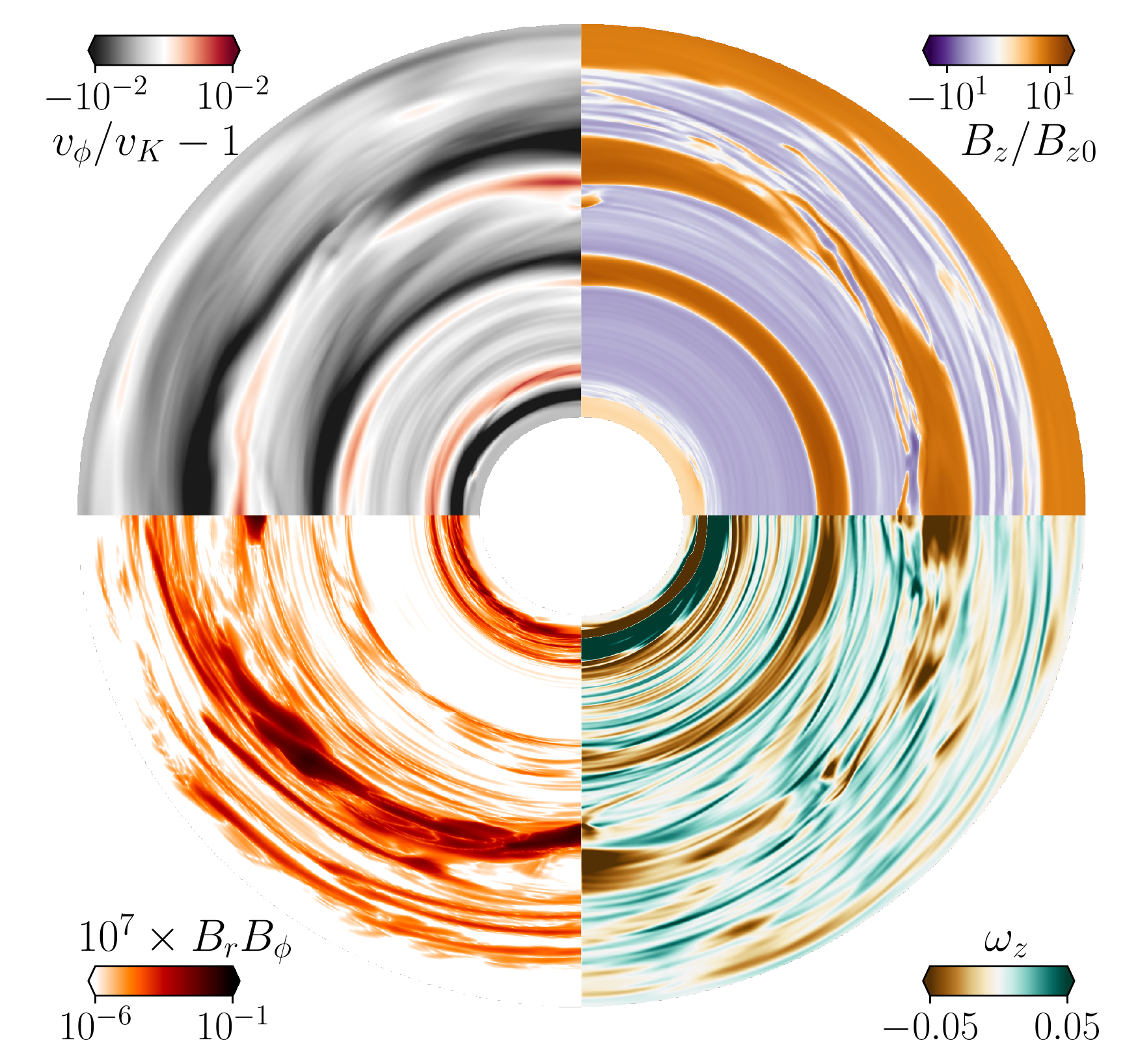} \\[3pt]
  \includegraphics[width=0.83\columnwidth]{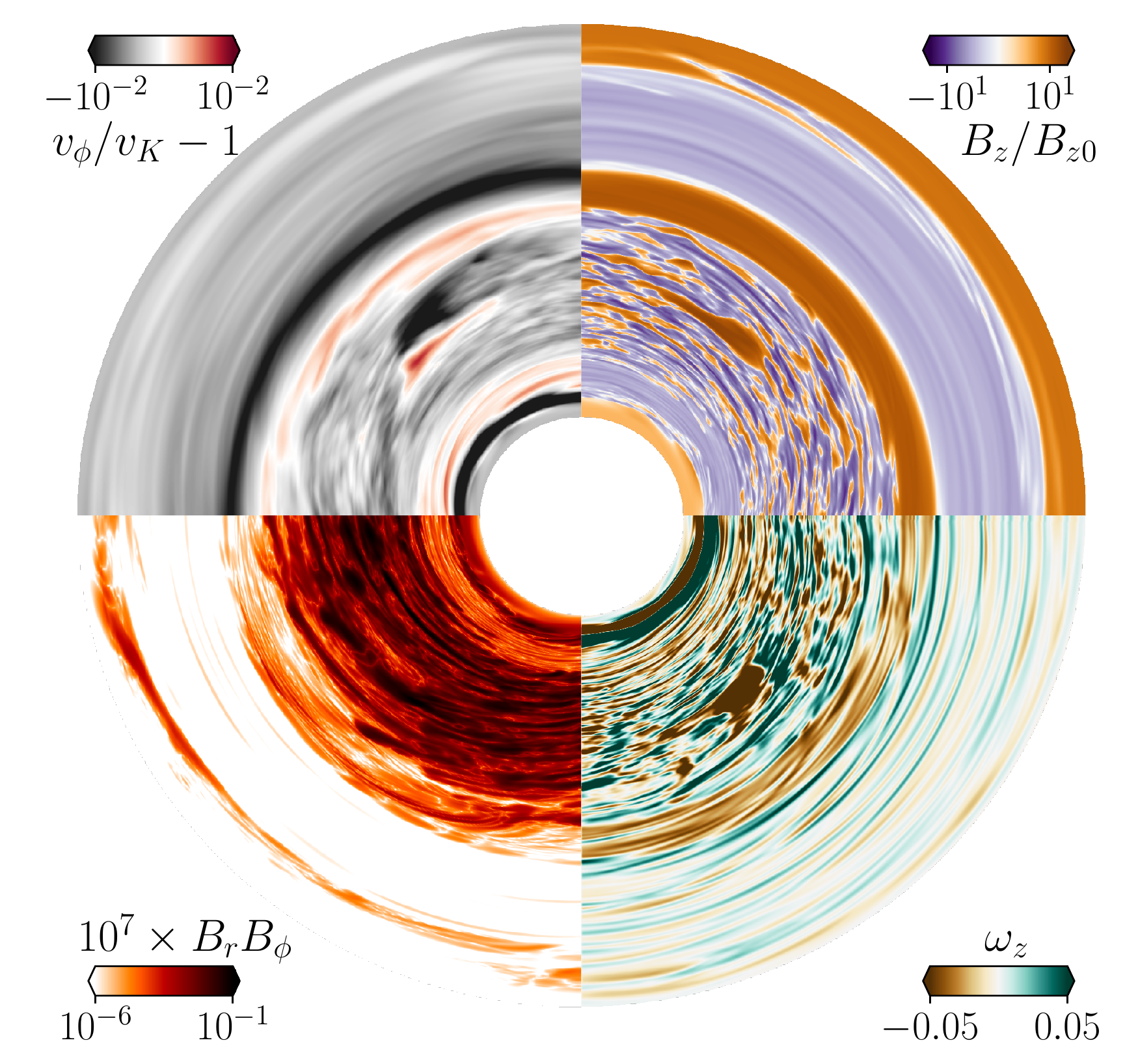}  \\

  \caption{Models with $\beta_{z}=10^3$, $5\times10^3$, and $10^4$ (top to bottom) at $t\eq250$, corresponding to models F3D-bz3-bp0, F3D-bz5.3-bp0, and F3D-bz4-bp0, as listed in \Table{runs}. In a clockwise sense, starting from the top right, sectors show: i) vertical field, ii) flow vorticity, iii) Maxwell stress, and iv) deviation from Keplerian rotation.}  \label{fig:nvf_aligned}
\end{figure}

The local regions of super-Keplerian velocity induced by the Hall-MHD self-organization seen in the previous section are stable in time and strong enough to halt the radial drift of the dust particles.
To see whether this trend holds for more realistic disk models, we study the conditions for dust trapping in a protoplanetary disk extended between $1-5\au$, using the disk model defined in \Sec{disk}, focusing on results obtained with \FTD.
The standard computational domain is $z\in[0,4h_0]$, $r\in [1,5]$, and $\phi\in [0,\pi/2]$, resolved with $32\times256 \times512$ grid cells, respectively.
We use a logarithmic spacing in the radial direction to maintain a constant radial resolution per gas pressure scale-height.

Vertical and azimuthal magnetic field components are defined via constant $\beta_z$ and $\beta_{\phi}$, respectively.
The aspect ratio is set to $h_0=0.05$, and the initial angular velocity of the gas is Keplerian.\footnote{The flow adjusts to the sub-Keplerian equilibrium during the first orbit.}
We add a white noise initial perturbation in the vertical and radial components of the velocity field in order to seed and grow the MRI in the ideal MHD limit.
After $t=40$ orbital periods, we switch on the Hall effect and Ohmic diffusion, and we further evolve the model until $t=300$.

The Hall diffusion coefficient, as introduced in \Sec{equations} --- see Eqns.~(\ref{eq:etaH}) and (\ref{eq:qH}) --- is given by
\begin{equation}
  \etaH(r) = q_{\rm H}\frac{h_0}{\sqrt{\rho_0}}\, \left(\frac{r}{r_0}\right)^{1+w}\, |\B|\,,
\end{equation}
with $w = 0.5$. We note that $\etaH(r)$ is a function of the \emph{initial} density, $\rho_0$, that is, the coefficient is held fixed during the evolution of the model.
We adopt an initial $\qH = 1$ at $r=r_0$, but we also run models with $\qH=2$, and $\qH=4$.

The Ohmic diffusion profile is given by $\etaO(r) = \eta_0\, r^{-1/2}$, and
we set $\eta_0 = 2\times10^{-6}$, which gives an initial $\Lambda_{\rm O} \simlt 1$ throughout the radial domain.
In \Figure{elsasser}, we show the initial Elsasser numbers computed with $\beta_z=10^4$ and $\beta_{\phi}=\infty$.

% --------------------------------------------------------------------------------

\begin{table}
  \caption{Overview of simulation runs}\label{tab:runs}
  \begin{tabular}{lccccc}
    \hline\hline
    & $\qH$ & $\beta_\phi$ & $\beta_z$ & $B_\phi$     & $B_z$        \\[-4pt]
    &       &              &           & $[{\rm mG}]$ & $[{\rm mG}]$ \\[+2pt]
    \hline
    \decimals
    F3D-bz1.4-bp0       & 1 & --        & $10^4$    & 0    &  48.1 \\[-4pt]
    F3D-bz1.4-bp0-2     & 2 & --        & $10^4$    & 0    &  48.1 \\[-4pt]
    F3D-bz1.4-bp0-4     & 4 & --        & $10^4$    & 0    &  48.1 \\
    F3D-bz5.3-bp0       & 1 & --        & $5\ee{3}$ & 0    & 68.0 \\[-4pt]
    F3D-bz5.3-bp0-2     & 2 & --        & $5\ee{3}$ & 0    & 68.0 \\[-4pt]
    F3D-bz5.3-bp0-4     & 4 & --        & $5\ee{3}$ & 0    & 68.0 \\
    F3D-bz5.3-bp5.3-2   & 2 & $5\ee{3}$ & $5\ee{3}$ & 68.0 & 68.0 \\[-4pt]
    F3D-bz5.3-bp50-2    & 2 & $50$      & $5\ee{3}$ & 680. & 68.0 \\[-4pt]
    F3D-bz5.3-bp5.3-4   & 4 & $5\ee{3}$ & $5\ee{3}$ & 68.0 & 68.0 \\[-4pt]
    F3D-bz5.3-bp50-4    & 4 & $50$      & $5\ee{3}$ & 680. & 68.0 \\
    F3D-bz1.3-bp0       & 1 & --        & $10^3$    & 0    & 152.1 \\[-4pt]
    F3D-bz5.3-bp0-fd    & 1 & --        & $5\ee{3}$ & 0    & 68.0 \\[-4pt]
    F3D-bz5.3-bp50-2-fd & 2 & $50$      & $5\ee{3}$ & 680. & 68.0 \\[+2pt]
    \hline
  \end{tabular}
  \medskip\newline \footnotesize{Values of $B_{\phi}$ and $B_z$ are given at $r=1\au$. The models with `f' have an azimuthal domain of $2\pi$ and the ones denoted with a `d' include dust.
  Model labels state parameter values used, such as `F3D-bz1.4-bp0-2', which for instance translates to: $\beta_z=10^4$, $\beta_\phi = \infty$, $q_{\rm H}=2$. }
\end{table}

% --------------------------------------------------------------------------------

\subsection{Self-organization for different vertical fields} % ---
\label{Bvertical}

We first discuss the runs with fixed $\beta_z$ and varying $\qH$, and then turn to results using three different initial $\beta_z$, while keeping $\qH$ fixed.

The parameters used for models F3D-bz1.4-bp0, F3D-bz1.4-bp0-2 and F3D-bz1.4-bp0-4 are listed in \Table{runs}, along with the other models.
For this first fiducial set of simulations, we fix $\beta_z=10^4$ and we set $\qH = 1$, $2$ and $4$, respectively.
Zonal flows are recovered in all these configurations.
The number of zonal flows that we observe increases as we increase $\qH$ from one band to four bands.

We consider two more models where we change the plasma-$\beta$ parameter to $\beta_z = 5\times10^3$ and $\beta_z = 10^3$, but we fix $\qH=1$.
We recover one zonal flow of vertical magnetic flux independently of the initial $\beta_{z}$, and in addition to the zonal flow, large-scale vortices show up when we decrease the $\beta_{z}$ parameter.

In \Figure{nvf_aligned}, we plot (in a clockwise sense), the vertical average of $B_z/B_0$ together with the the vertical component of the vorticity $\omega_z \equiv (\nabla \times \vg)_z$, the Maxwell stress and the deviation from the Keplerian rotation profile, $v_{\phi}/v_{\rm K} -1$.
The plasma-$\beta$ parameter increases from $\beta_{z} = 10^3$ (top), followed by $5\times10^3$ (middle) and finally $10^4$ (bottom panel).
For the case of $\beta_{z}\eq 10^4$, a concentration of substantial Maxwell stresses is located in the region between the zonal flow and the inner radius.
Adjacent to the zonal flow, the Maxwell stress decreases considerably.
For the other two cases, it is possible to recognize regions with a lower amount of stress across the vortices.
The Maxwell stress decreases as well between the zonal flows, and turbulent fluctuations persist near the regions with enhanced vertical magnetic field.

In order to confirm the ability of these models to affect the dust evolution, we show, in the same figure, the vertical average of the deviation from Keplerian rotation, $v_{\phi}/v_{\rm K}-1$.
In contrast to the setup described in section \Sec{self}, the initial conditions here impose an equilibrium rotation profile with sub-Keplerian velocity.
This implies that stronger local pressure gradients are need to locally reach super-rotation.
However, for all the models, we find super-Keplerian regions with a velocity deviation of 1\%, i.e., $v_{\phi}/v_{\rm K} -1 \simeq 0.01$. These regions are stable for $t\simgt100$.

\begin{figure*}
        \centering
        \includegraphics[width=\textwidth]{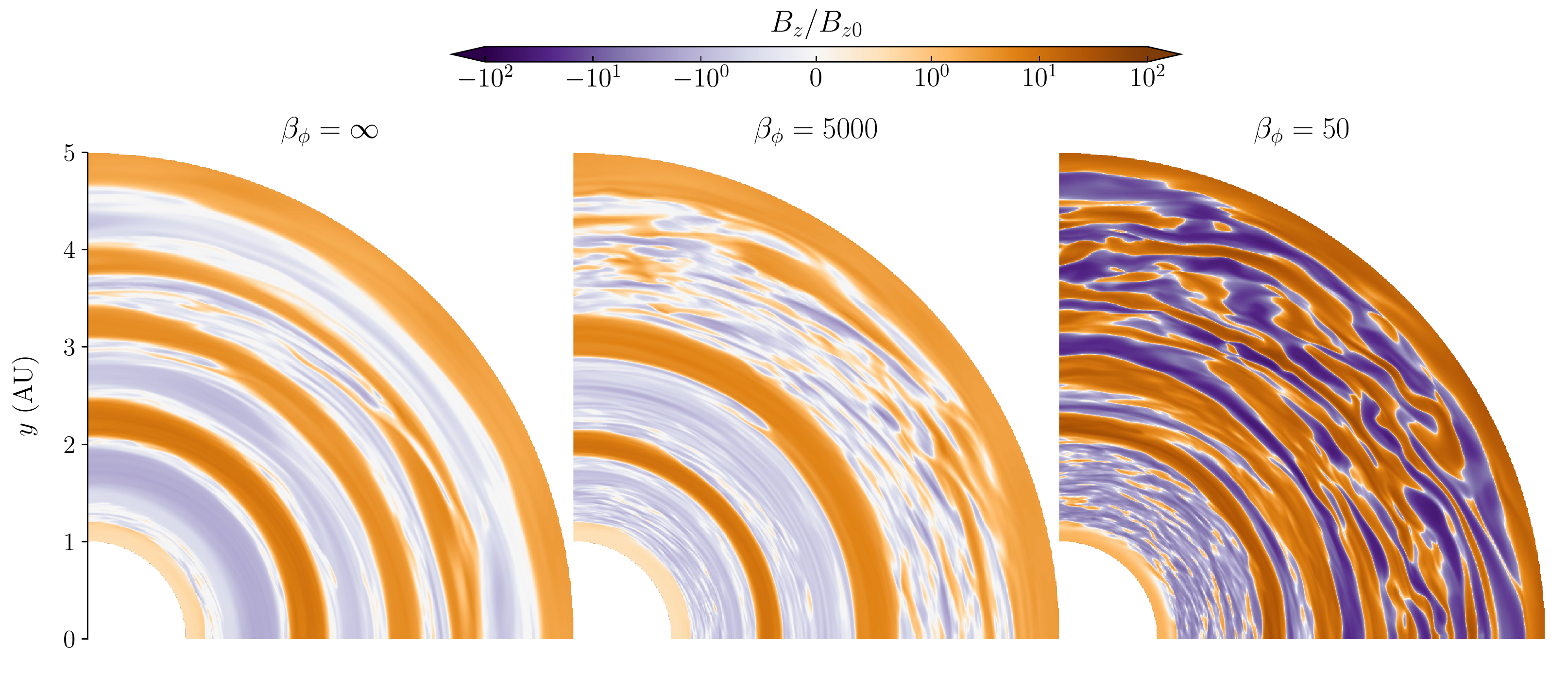} \\
        \caption{Vertical magnetic field, $\mn{B_z}$ at $t\eq200$ for runs F3D-bz5.3-bp0-2 (left) ,F3D-bz5.3-bp5.3-2 (center) and F3D-bz5.3-bp50-2 (right), with initial values $\beta_{z} = 5\ee{3}$ and $\qH = 2$. Note that the color scale is \emph{signed} logarithmic.} \label{fig:azi}
\end{figure*}

\subsubsection{Generalized flow vorticity} % ---

In the absence of dissipative terms --- that is, viscosity and Ohmic diffusion --- the magneto-vorticity $\omegaM$ satisfies a local conservation equation  \citep{Polygiannakis2001,Pandey2008,Kunz2013a} if the fluid is incompressible.
This quantity is defined as
\begin{equation}
  \omegaM \equiv \nabla \times \vg + \omega_{\rm H}\, \bb\,,
\end{equation}
where $\omega_{\rm H} \equiv |B| n_e e/\rho$ is the Hall frequency. In the limit of small Mach numbers, $\omegaM$ can still be regarded as an approximately conserved quantity.

The local conservation of $\omegaM$ implies that a concentration of vertical flux has to be anti-correlated with a low vorticity region.
Despite the fact that we include Ohmic diffusion, and the flow is moderately compressible, we do expect to find regions where $\omega_z$ decreases simultaneously when concentrations of the magnetic flux appear.
To test this hypothesis, we compute $\omega_z$ using the vertical average of the perturbed velocity field.
We plot the vorticity in the bottom right sector of the disks shown in \Fig{nvf_aligned}. By comparing the top left and bottom right sectors, respectively, one can see that negative vorticity regions coincide with the places where the vertical magnetic flux is accumulated --- a trend that becomes more prominent for stronger fields, i.e., as $\beta_{z}$ decreases.

\subsection{Inclusion of a net azimuthal magnetic flux} \label{sec:Bazimuthal}

When including a weak azimuthal net flux, that is, $\beta_{\phi} \sim \beta_z$, \citet{Bethune2016a} found that the global picture of self-organization remains intact, despite the presence of a slightly enhanced turbulent activity.
We study the effect of a net azimuthal flux on the dynamics by first including an initial $\beta_{\phi}$ equal to $\beta_{z}=5\ee{3}$ (cf. row 7 in \Tab{runs}), and then increasing the former by a factor of a hundred, that is, $\beta_{\phi}=50$ and $\beta_{z}=5\ee{3}$ (cf. row 8 in \Tab{runs}).

\begin{figure}
  \centering
  \includegraphics[width=\columnwidth]{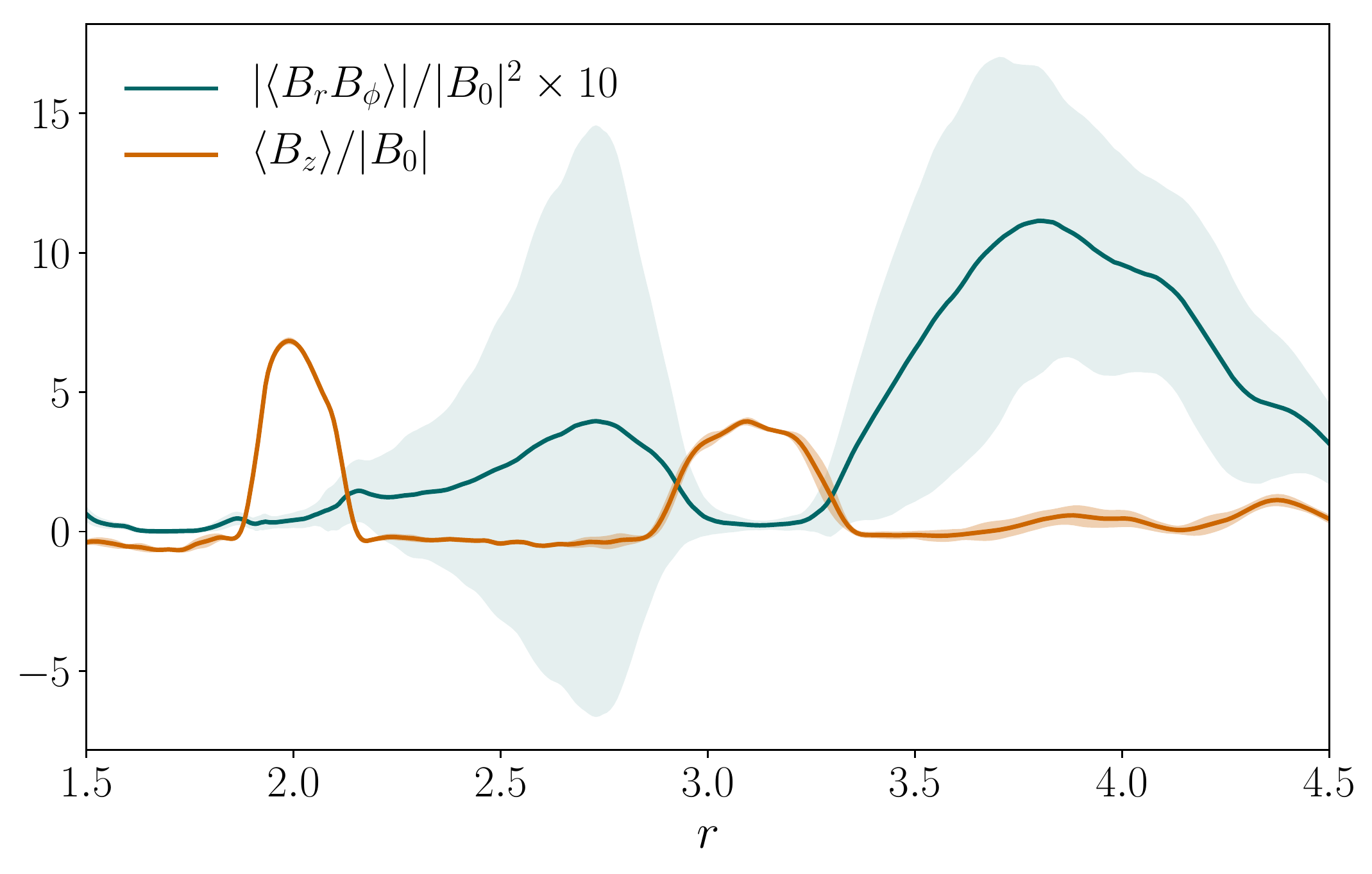} \\[1pt]
  (a) --- weak azimuthal field, $\beta_{\phi} = 5000$         \\
  \includegraphics[width=\columnwidth]{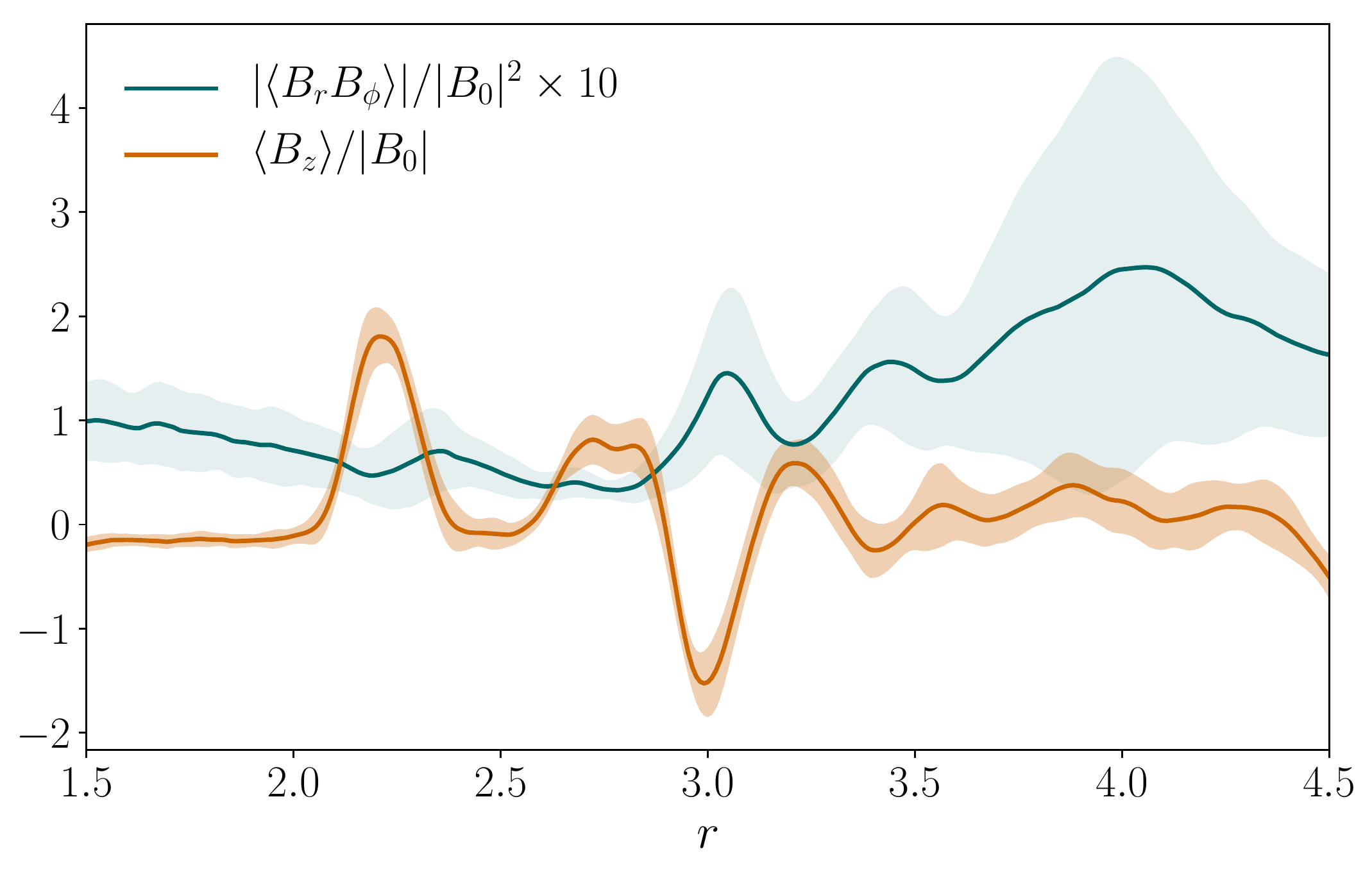} \\[1pt]
  (b) --- strong azimuthal field, $\beta_{\phi} = 50$
  \caption{Radial profiles of $\mn{B_z}$ and the Maxwell stress. The shades indicate the standard deviation of the time average computed between $t=100$ and $t=200$.} \label{fig:azi_stress}
\end{figure}

In \Figure{azi}, we show  the vertical magnetic field after 200 orbits for $\beta_z = 5\ee{3}$, $\qH = 2$ and different values of $\beta_\phi$.
In panels (a) and (b) of \Fig{azi_stress}, we plot the mean radial profile, between $t=100$ -- $200$, of the vertical magnetic field and the Maxwell stress.
When $\beta_{\phi} \simeq \beta_z $, we recover two zonal flows with adjacent regions of turbulent fluctuations.
This is in agreement with the field morphology found with $\beta_{\phi}=\infty$, where we obtained three zonal flows.

The disk dynamics, however, evolve differently when $\beta_{\phi} = 50$ is used.
Panel (b) of \Fig{azi_stress} shows a prominent region of Maxwell stress in the outer disk, where large-scale turbulent perturbations dominate the dynamics.
The self-organization mechanism is not clearly distinguishable despite the fact that two adjacent zonal flows are recognized.
These ring-like structures are radially thinner than the zonal flows obtained with a weak azimuthal field, but are also stable in time.
Furthermore, they are correlated with super-Keplerian velocity regions, so again we obtain azimuthally large-scale regions where the radial drift of the dust is slowed down.
\Figure{azi} conveys a clear trend, where the inclusion of a net azimuthal field prevents the self-organization, but azimuthally large-scale flux concentrations are still possible to obtain.

We have inspected the Reynolds and Maxwell components associated with the $\alpha$-parameter introduced in \Sec{previous} --- see \Eqn{alpha}.
The contribution of the Reynolds stress is below 1\% of the total stress, except in the regions where the zonal flows are sustained.
In these regions, the Maxwell stress decreases and the local perturbations of the magnetic field are comparable to the velocity turbulent fluctuations.
When no azimuthal flux is considered, the Maxwell contribution to the $\alpha$-parameter is, on average, $\sim10^{-4}$.
Over the zonal flows it drops to $\sim10^{-6}$, whereas the Reynolds contribution has a radially constant mean of $\sim10^{-6}$.
When $\beta_{\phi} = 50$, the ratio between the Maxwell and Reynolds contributions is comparable to the case with zero net azimuthal flux, but the amplitude of the perturbations is higher by a factor of $\sim100$ for both.

The decreased appearance of self-organized structures as a consequence of including an azimuthal field, $\B_{\phi}$, much stronger than $\B_z$ is in agreement with the previous work of \citet{Lesur2014}. They included vertical stratification in a shearing box model, resulting in a strong azimuthal field generated by the Hall effect.
Furthermore, they found that the mean azimuthal flux can be 200 times larger than the mean vertical flux, even if the initial azimuthal flux is negligible.
A similar result was obtained by \citet{Bethune2017} using a spherical global disk configuration with a magnetized wind in a regime where $\beta_z = 10^2$.
The authors showed that ambipolar diffusion favors the accumulation of vertical magnetic field into zonal flows. Despite the fact that the Hall effect is negligible (compared with the ambipolar diffusion), it can act against the self-organization if the wind can drive a magnetic stress in regions of strong field.

Even though we do not include vertical stratification, we observe that the inclusion of a strong azimuthal flux can alter the dynamics, inhibiting the organization of the zonal flows between channels of strong Maxwell stress and creating a more turbulent flow.
However, a sufficient strong Hall diffusion ($\qH \simgt 2$) leads to azimuthally large-scale structures of the vertical magnetic field that generate super-Keplerian velocity regions which are stable for more than 100 orbits.
This potentially implies that, even without a clear appearance of self-organized  features, it is possible to find flow regions where the dust drift slows down or even halts.

\subsubsection{Spectral energy distributions} % ---

In order to establish a more quantitative comparison between these cases, we compute the spectral energy distribution of the vertical and azimuthal field between radius $r\eq2$ and $4$ for models F3D-bz5.3-bp0-2, -bp5.3-2 and -bp50-2.

\begin{figure}
  \centering
  \includegraphics[width=0.9\columnwidth]{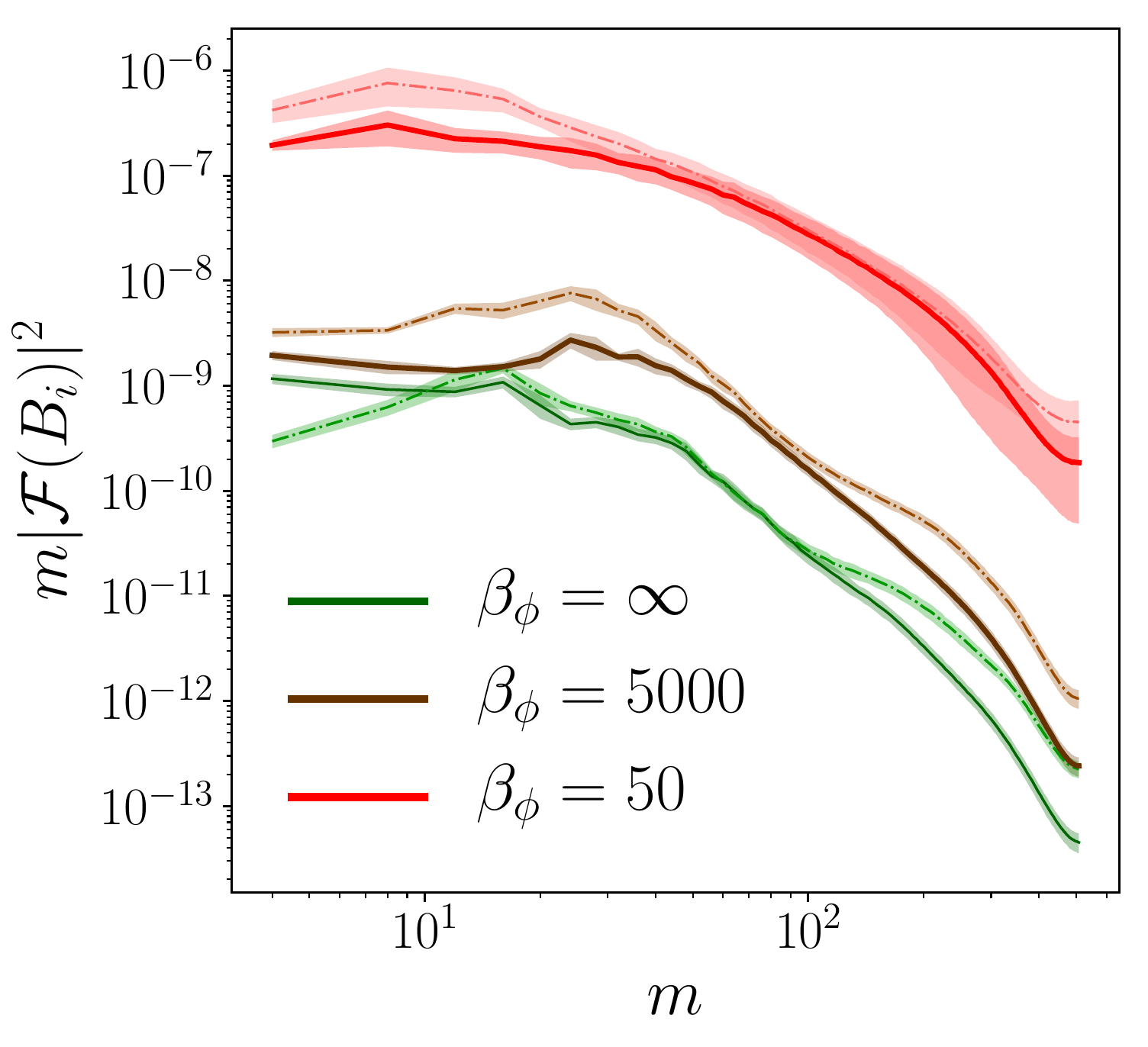}
  \caption{Spectral energy distribution, $m\,|\mathcal{F}(B_s)|^2$, for $B_z$ (solid lines) and $B_{\phi}$ (dotted lines) for weak and strong azimuthal net flux (averaged between $t=190$ and $t=210$).} \label{fig:bp_spectrum}
\end{figure}

In \Figure{bp_spectrum}, we plot the spectral energy distribution, $m\,|\mathcal{F}(B_i)|^2$, for the azimuthal and vertical field as a function of the azimuthal mode number, $m$, where the smallest mode is $m = 4$.
For the runs with $\beta_{\phi}=5\times10^3$, the maximum of the distribution is located at large-scale modes -- but the peak is clearly reached inside the azimuthal domain $\pi/2$ with an almost flat distribution at lower $m$.
Increasing the azimuthal net-flux to $\beta_{\phi}=50$ shows a different distribution.
The maximum is reached around $m\sim 4$ for the components $B_z$ and $B_{\phi}$.
This implies that a full $2\pi$ domain might be needed in order to correctly capture the maximum of the energy distribution.
The energy grows to larger length scales, which can be recognized in \Fig{azi}, where large-scale field perturbations dominate the vertical field structure.

\begin{figure}
  \centering
  \includegraphics[width=0.4\textwidth]{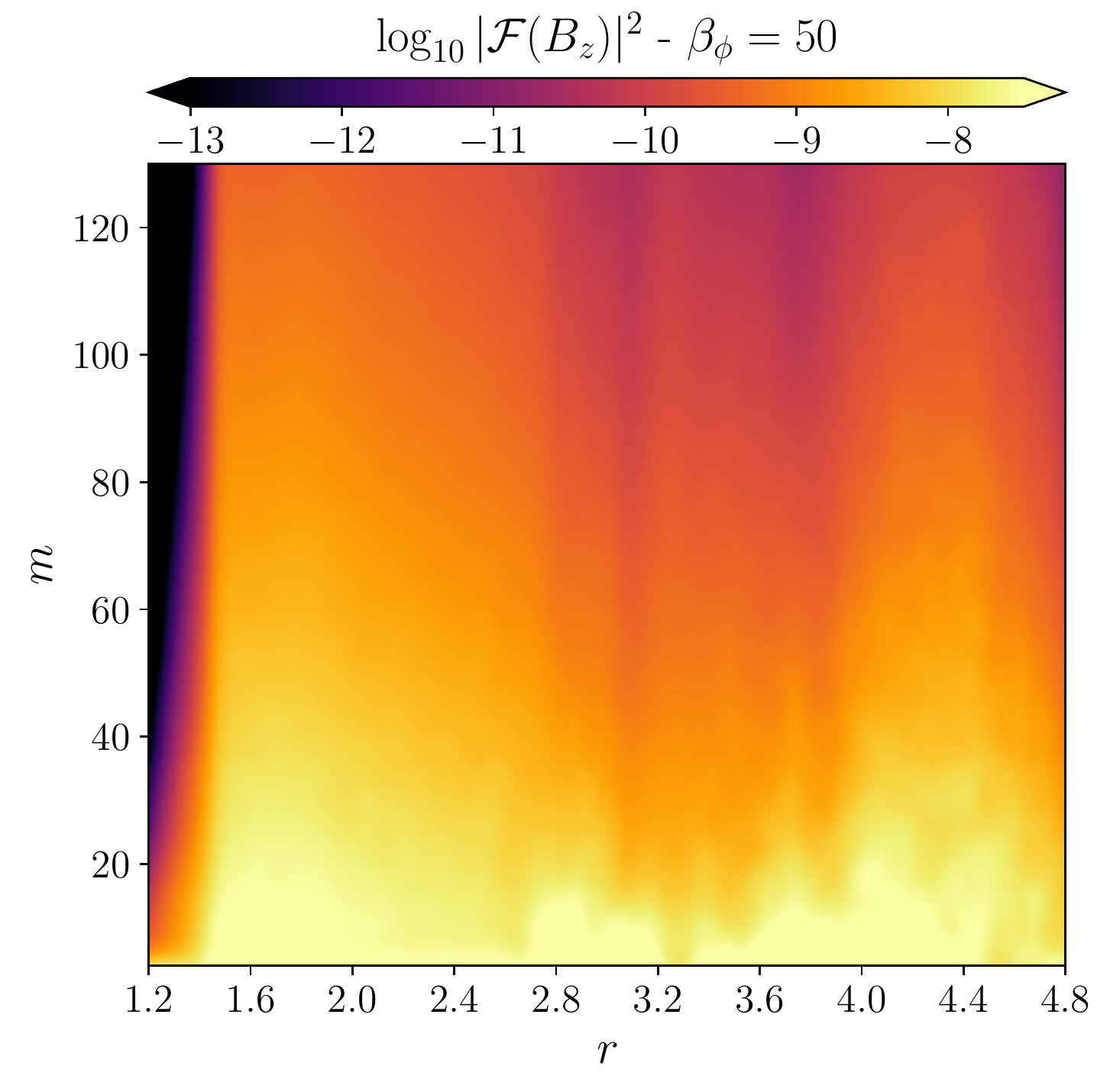} \\
  \includegraphics[width=0.4\textwidth]{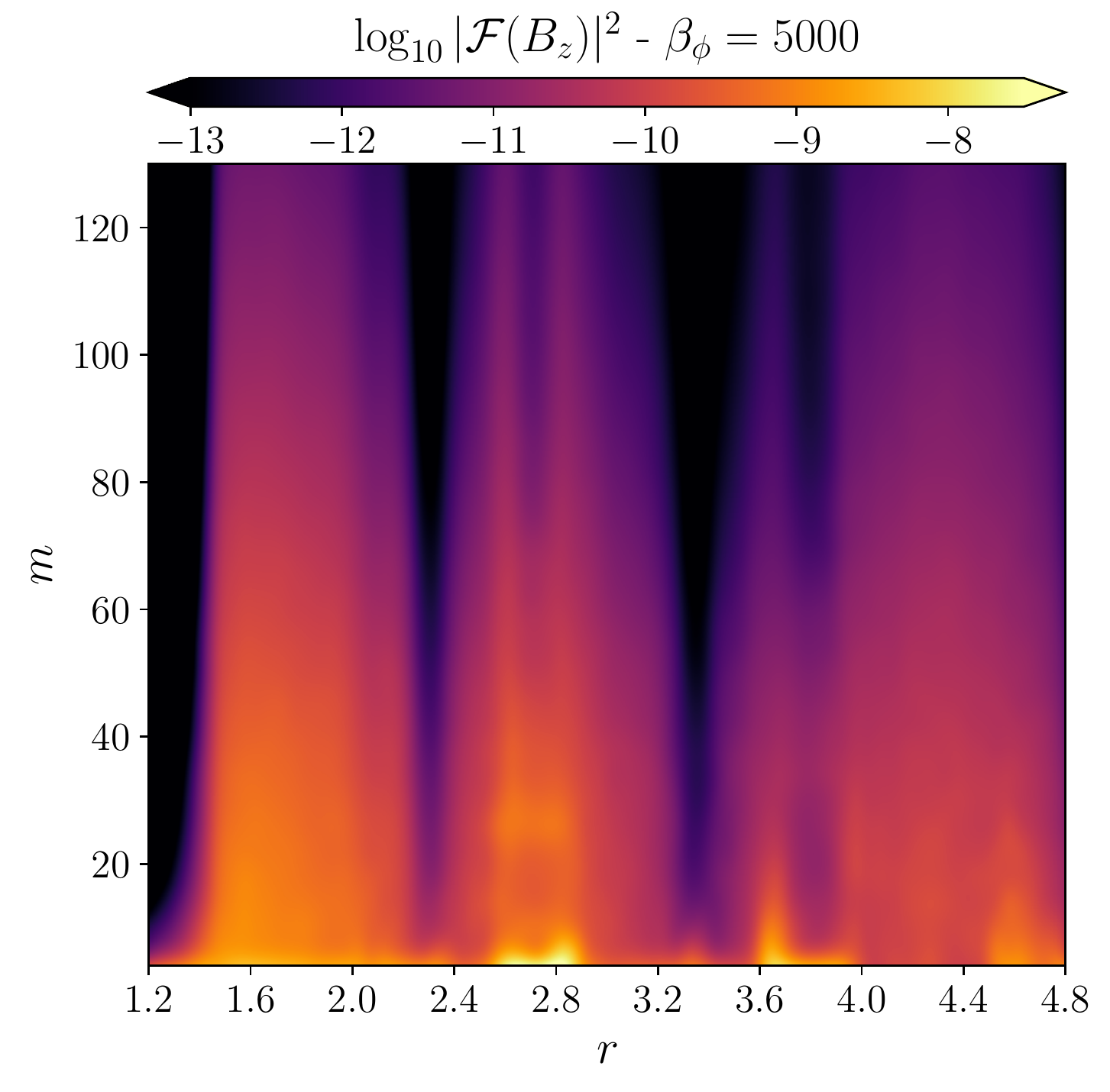} \\
  \caption{ Spectral energy distribution $|\mathcal{F}(B_z)|^2$, for 80 radial uniform bins, averaged between $t=190,210$. \emph{Top:} model  F3D-bz3.5-bp50 with $\beta_{\phi}=50$. \emph{Bottom:} model -bp5.3 with $\beta_{\phi}=5000$.} \label{fig:bp_spectrum2D}
\end{figure}

The 1D spectrum in \Fig{bp_spectrum} allows us to distinguish the trends of the different spectral energy distributions, but it might be affected by the specific choice of the spatial domain.
To confirm the differences and obtain a clearer picture, we show in \Fig{bp_spectrum2D} the spectral energy distribution for $B_z$ as a function of radius.
To this end, we divide the radial domain into 80 uniformly-spaced bins and compute the azimuthal spectrum for each annulus.
The bottom panel shows the case with $\beta_{\phi}=5\ee{3}$, where it is possible to recognize the location of the two zonal flows.
In these regions, the energy is mainly concentrated in low-$m$ modes,
and drops as we move outward in the disk. Perturbations are restricted to modes $m\simlt40$.
The upper panel shows the spectral distribution for $\beta_{\phi}=50$,
where the energy has a more homogeneous distribution and extends to modes with $m\simgt100$.

% ------------------------------------------------------------------------------

\section{Dynamics of embedded dust grains} \label{sec:dust} % ---

\begin{figure*}
  \includegraphics[width=\textwidth]{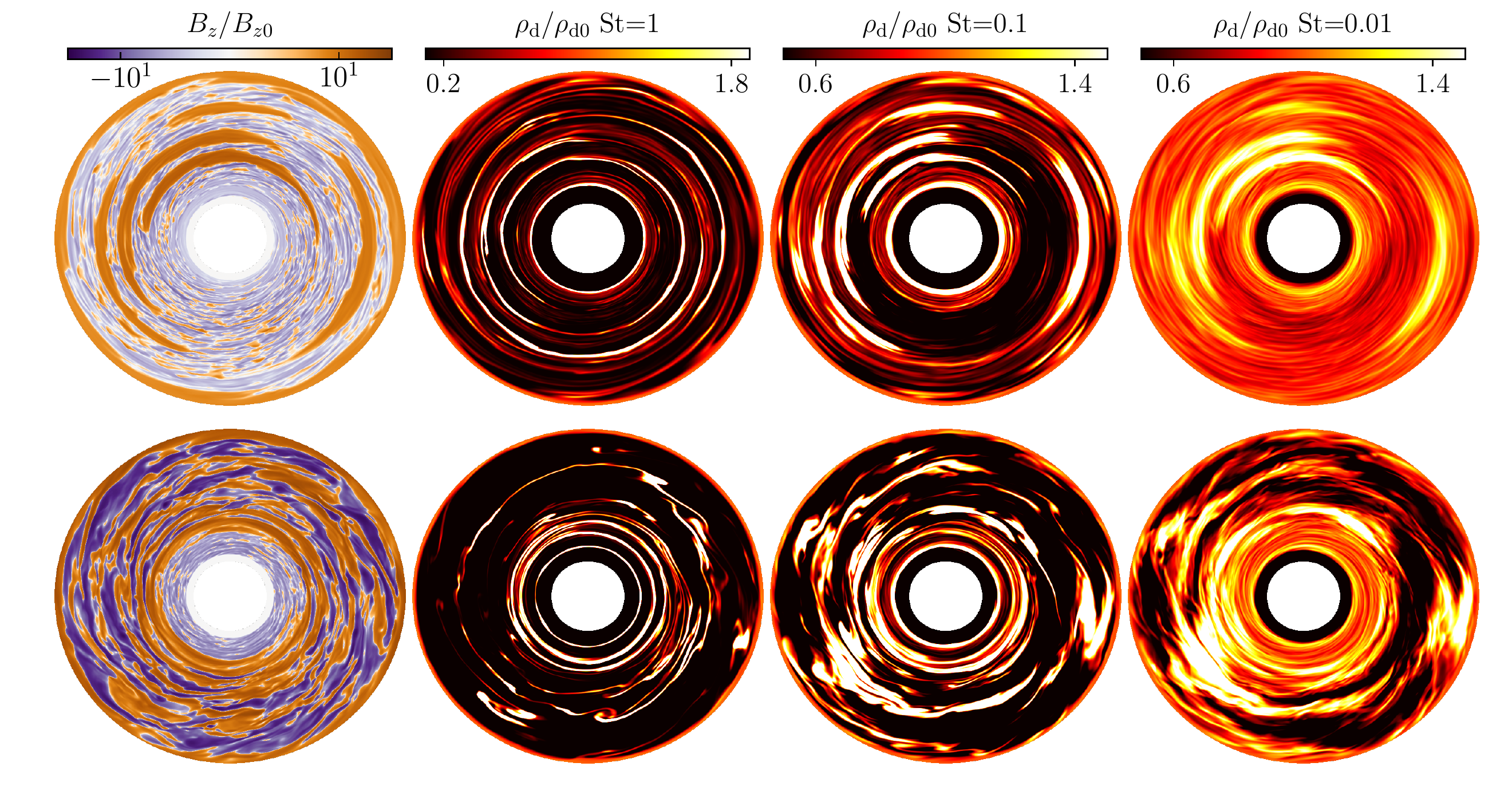}
  \caption{Vertical average for $B_z/B_{z0}$ (leftmost panel) and dust density contrast $\rhod/\rho_{\rm d0}$ (remaining three panels) for $t=250$. The dust density contrast is shown in order of decreasing Stokes number of 1, $0.1$, and $0.01$, respectively.
  The upper row corresponds to model F3D-bz5.3-bp0-fd, without an azimuthal field, and the bottom row corresponds to the model F3D-bz5.3-bp50-fd, with a dominant azimuthal field.}
  \label{fig:dust}
\end{figure*}

In view of the azimuthally large-scale morphology of the flux concentrations, and accommodating for the fact that the spectral energy distribution is dominated by low azimuthal wavenumbers, we now adopt a full-disk domain, that is, $L_\phi = 2\pi$ with 1024 uniform spaced cells.
This is motivated by the possibility that the zonal bands that we observe in a disk with an azimuthal domain of $L_\phi=\pi/2$ might develop into vortices in a $L_\phi=2\pi$ domain.
This notion was already appreciated by \citet{Bethune2016a} and, as we will show, it is relevant to the models that we present here.

We now turn to the models F3D-bz5.3-bp50-fd and -bp0-fd to study the dust evolution with three different Stokes number, $\St=0.01$, 0.1 and 1 (see \Sec{equations} for definitions).
Models F3D-bz5.3-bp0-fd and -bp50-fd-2 have the same initial conditions as F3D-bz5.3-bp0 and -bp50-2, respectively.
The dust is considered as a pressureless fluid with an initial density $\rhod = 0.01 \rhog$, and we, moreover, neglect its feedback onto the gas.
We enable the drag force term acting on the dust at $t\eq180$ inner orbits and evolve the system until we reach $t\eq260$.

In \Figure{dust}, we show the vertically-averaged dust-density contrast, $\rhod/\rhodz$, for the three different Stokes numbers.  We moreover plot the vertical average of $B_z/B_{z0}$, for which we expect to find flux concentrations that coincide with locations of dust accumulations.

For the $L_\phi\eq2\pi$ models, large-scale vortices appear, which is in contrast to the reduced $L_{\phi}\eq\pi/2$ model, where one zonal flow and two vortices were obtained (see the middle panel of \Fig{nvf_aligned}).
In agreement with the model F3D-bz5.3-bp0, these vortices induce patches of strong super-Keplerian velocities.
These regions all show an enhancement of the dust density for the three different Stokes numbers that we have studied here.
After a period of $t=80$ inner orbits, the different grain sizes already display different levels of local concentration, which simply reflects the segregation according to the particles' Stokes number.
The dust species with $\St=0.1$ and $\St=0.01$ are more coupled to the gas than the species with $\St=1$.
In particular, for $\St=0.01$, we obtain density variations smaller than 10\% with respect to the gas.
The reason why we do not see similar dust concentrations in all three components, can be understood in terms of their different radial drift time scales.
Let us consider a 2D equilibrium velocity dust profile \citep{Nakagawa, Takeuchi2002} and assume that the radial velocity of the gas can be neglected.
This assumption is an approximation that not fully applies to our simulations, but nevertheless, it allows us to estimate the drift time scale $\tau$ for each Stokes number, yielding
\begin{equation}
  \frac{\tau}{T_0} = \frac{1}{9\pi}\frac{\St + \St^{-1}}{h_0^2} \bigg[ \left(\frac{r}{r_0}\right)^{\!3/2} \!- 1  \,\bigg] \,,
\end{equation}
where $T_0 = 2\pi/\Omega_0$.
If we consider a trap located at $r=3$, the drift time scale is $\tau/T_0 \simeq 118 $ inner orbits for unity Stokes number -- meanwhile, for $\St = 0.01$, it is around $\tau/T_0 \simeq 6\ee{3}$, so we need to integrate 6000 orbits for $\St=0.01$ in order to see a similar dust enhancement as that of the species with $\St=1$.

\begin{figure*}
  \centering
  \includegraphics[height=1.4\columnwidth]{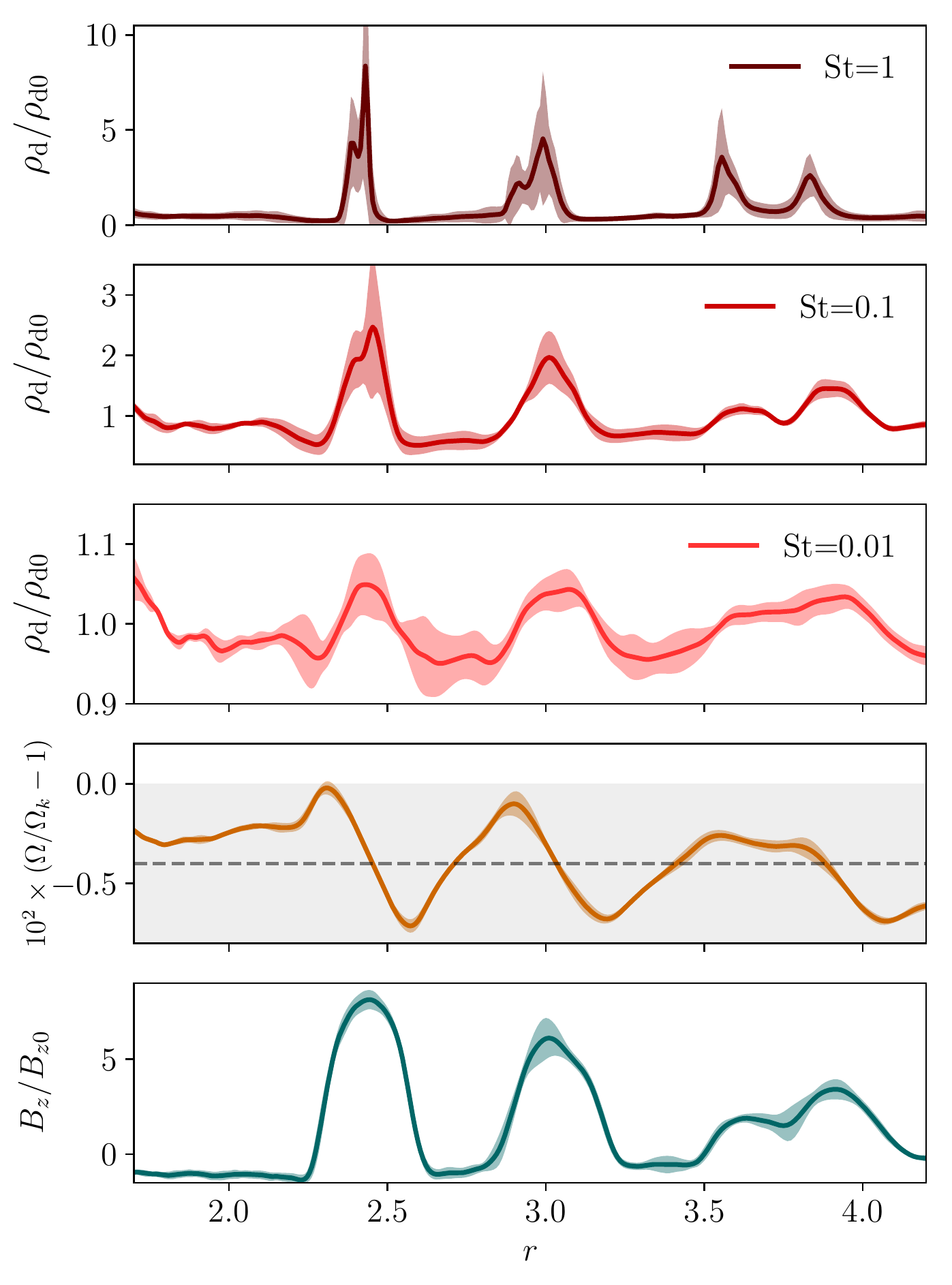}\hfill
  \includegraphics[height=1.4\columnwidth]{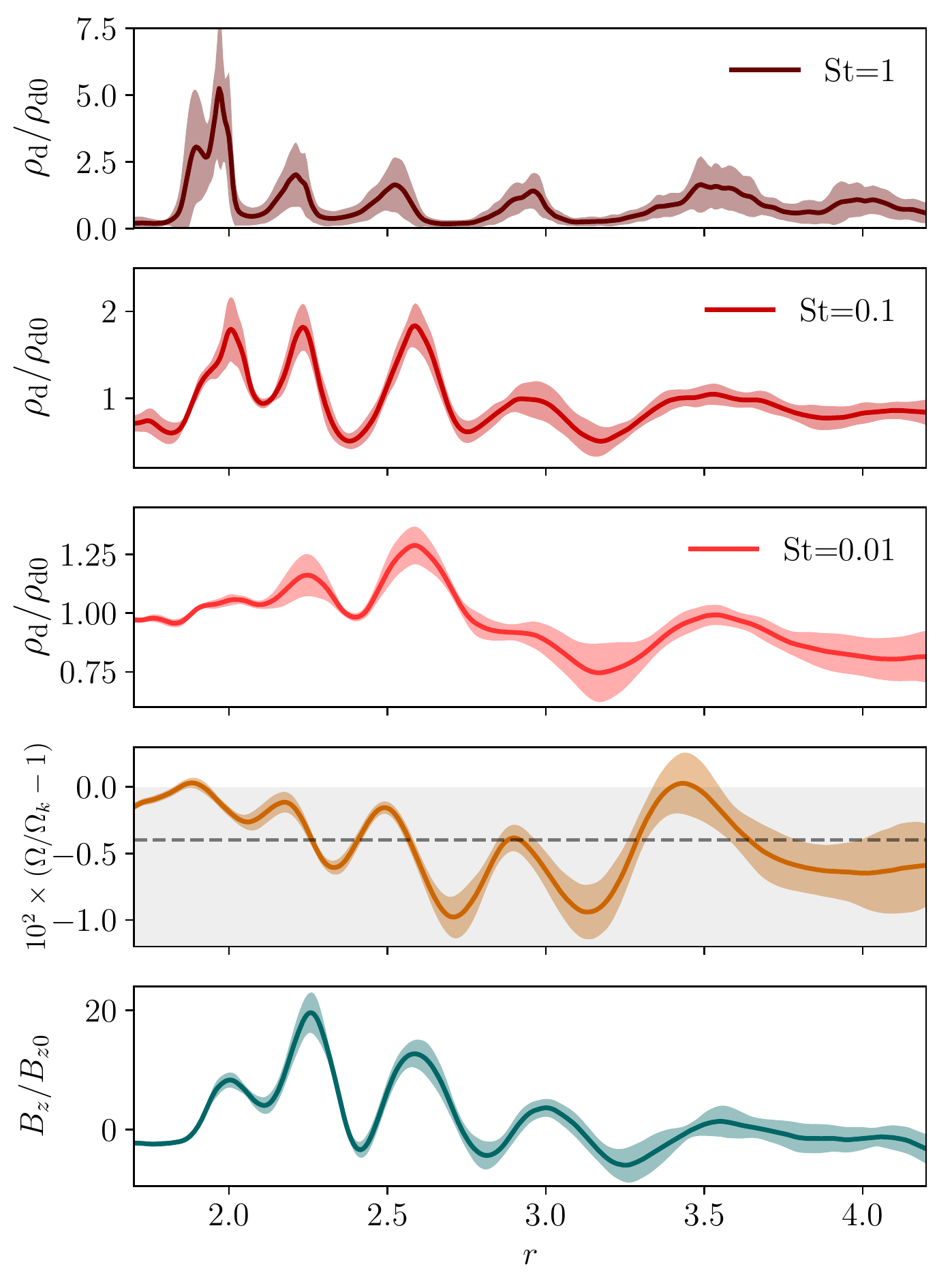}\\[-6pt]
  \caption{Mean radial profiles (averaged between $t=190 - 260$) for models F3D-bz5.3-bp0-fd, with zero azimuthal flux (left panel), and  F3D-bz5.3-bp50-fd, with a strong azimuthal flux (right panel). Dashed lines indicate the (sub-Keplerian) equilibrium velocity profile.}\label{fig:1Ddust}
  \medskip
\end{figure*}

In \Figure{1Ddust}, we plot azimuthally averaged radial profiles of the dust-density ratios $\rhod/\rhodz$, along with the deviation, $\Omega/\Omega_K -1 $, from Keplerian rotation, and the vertical magnetic field enhancement, $B_z/B_{z0}$.
Despite not being axisymmetric, the vortical structures are sufficiently large in their azimuthal extent that averaging the velocity deviation in the toroidal direction will enhance regions were the dust accumulations are stable in time.
Taking azimuthal averages moreover brings out regions where the average velocity field  becomes almost super-Keplerian over significant intervals of time -- even though the super-Keplerian rotation of the vortices does get washed-out by taking the mean.
This allows us to identify the radial locations of strong magnetic flux concentrations and recognize that these are the regions of dust enhancement.
For Stokes number $\St=1$, in some locations, the density contrast reaches a peak of $\rhod/\rhodz \simeq 20$.
For longer times, it might reach even higher values, and the feedback onto the gas can no longer be neglected.

In agreement with the discussion in section \Sec{Bazimuthal}, the inclusion of a strong azimuthal field induces large-scale concentrations of the vertical magnetic field, which in this particular case can be axisymmetric.
These regions are generally stable in time and induce super-Keplerian flows.
The profiles shown in the right panel of \Fig{1Ddust} highlight the coincidence of these locations with the dust enhancements.

% ------------------------------------------------------------------------------

\section{Conclusion} \label{sec:conclusion}

We have demonstrated that our implementations of Hall-MHD in \NIR and \FTD are suitable for studying problems in the context of protoplanetary disks subject to the non-ideal MHD effects.
The inclusion of an artificial resistivity is only necessary when the Hall effect strongly dominates in the induction equation by a factor more than $10^2$, with respect to the ideal-MHD induction term.
In all the models explored in this strong Hall regime, around 5\% of the active domain has an artificial magnetic Reynolds number $1<\Rma<5$, so we conclude that the dissipation likely has a minimal impact on the dynamics and that it overall does not affect the results that we have obtained.

A central new result of our investigation is that the self-organization is sustained in a disk with radial stratification. This is also true if the azimuthal flux is comparable to the vertical flux for a plasma-$\beta$~parameter in the range between $\beta_{z} = 10^4 - 10^3$.
Simulations where the azimuthal domain is reduced to $L_{\phi} = \pi/2$ show axisymmetric zonal flows and vortices, meanwhile we are only able to recognize large-scale vortical features if the domain is $L_{\phi}\eq2\pi$.

We confirm the finding that the zonal flows (and vortices) are confined between regions of strong Maxwell stress --- in agreement with the mechanism described by \citet{Kunz2013a}.
Because of the low flow compressibility and the inclusion of small Ohmic dissipation, the vortices and zonal flows are anti-correlated with low vorticity regions, which is expected when the magneto-vorticity is locally conserved.

When the initial model has a strong azimuthal net flux, i.e., $\beta_{\phi} \simeq 10^{-2}\tms \beta_{z}$, azimuthally large-scale concentrations of vertical magnetic flux are still possible to obtain.
However, the picture of field confinement between regions of enhanced Maxwell stress is no longer clearly identified in these simulations.
The azimuthal component of the magnetic field strongly dominates the dynamics and the self-organized zonal flows are harder to recover, which is in agreement with the previous results by \citet{Lesur2014} and \citet{Bethune2017}.
The inclusion of a net-azimuthal flux does not alter the evolution of the zonal flows when $\beta_{\phi}\sim \beta_z$ \citep[also cf.][]{Bethune2016a}.
The spectral energy distribution increases as we move towards shorter azimuthal wavenumber, $m \simeq 4$.
When $\beta_{\phi}=5\ee{3}$, the distribution shows a maximum followed by a flat profile at the lower azimuthal modes.
Increasing $\beta_{\phi}$ leads to a spectrum that has a peak at intermediate azimuthal wavenumbers.
These energy distributions have been obtained using a disk with a reduced azimuthal extent of $\pi/2$.
The fact that the maxima are located close to the lowest azimuthal wavenumbers implies that a full $2\pi$ domain is advisable for studying the global evolution of the zonal flows.

By including the evolution of pressureless dust fluids in the ensuing Hall-MHD turbulence, we demonstrate that quasi-axisymmetric dust enhancements can be obtained for the range of Stokes numbers explored -- even in the absence of prominent flow features, such as the result of Hall-effect induced self-organization.
There appears to be a difference in character regarding the precise nature of the ensuing dust traps, however.
In models \emph{without} azimuthal net flux,  dust enhancements are typically located at the position of vortices themselves, which agrees well with the notion of vortices being able to trap dust \citep{1997Icar..128..213K}.
In contrast to this, in models with significant azimuthal magnetic flux ($\beta_\phi\simeq 50$), the dust accumulations appear to coincide with local concentrations of the vertical magnetic flux.

For particles with Stokes number $\St\eq1$, peaks in the dust concentration with $\rho_{d}/\rho_{d0} \simeq 20$ are reached during the integration time of a few hundred inner orbits that we have adopted here, implying that the feedback onto the gas might have to be considered when pursuing longer integration times.
The drift time scale of the $\St=0.1$ and $\St=0.01$ particles can be estimated to become between one and two orders of magnitude higher than that for $\St=1$. As a result, lower dust enhancement factors were obtained for these two dust species, which we, at least partially, attribute to the insufficient evolution time covered by our current simulations.

In conclusion, we have demonstrated the ability of our numerical implementations to deal with the MHD turbulence created by the Hall-MRI in the regime of a dominant Hall effect, and were able to study its consequences for the local accumulation and segregation of intermediate-Stokes number dust grains.
The presented numerical tools will enable us to further explore the phenomenon of self-organization in the context of Hall-MHD turbulence, as well as its potential relevance in explaining the axisymmetric and non-axisymmetric substructure observed in nearby protoplanetary disks in the (sub-)mm waveband.

% ------------------------------------------------------------------------------
\software{ NIRVANA-III
        \citep{2004JCoPh.196..393Z,2011JCoPh.230.1035Z,2016A&A...586A..82Z}, FARGO3D \citep{Benitez-Llambay2016}, IPython \citep{numpy}, Matplotlib \citep{matplt} }
% ------------------------------------------------------------------------------
\acknowledgments

We thank Philipp Weber, who provided useful comments as well as the referee for a valuable report.
This project has received funding from the European Union's Horizon 2020 research and innovation programme under grant agreement No 748544 (PBLL).
The research leading to these results has received funding from the European Research Council under the European Union's Horizon 2020 research and innovation programme (grant agreement No 638596) (OG).
The research leading to these results has received funding from the European Research Council under the European Union's Seventh Framework programme (FP/2007-2013) under ERC grant agreement No 306614 (MEP).
This research was supported in part by the National Science Foundation under Grant No. NSF PHY17-48958.
This research was supported by the Munich Institute for Astro- and Particle Physics (MIAPP) of the DFG cluster of excellence ``Origin and Structure of the Universe''.
This work used a modified version of the \NIR code based on v3.5 developed by Udo Ziegler at the Leibniz Institute for Astrophysics, Potsdam (AIP).
We acknowledge that the results of this research have been achieved using the PRACE Research Infrastructure resource MareNostrum-4 based in Spain at the Barcelona Supercomputing Center (BSC).
Computations were performed on the \texttt{astro\_gpu} partition of the Steno cluster at the University of Copenhagen HPC center.

% ------------------------------------------------------------------------------

\appendix

% -------------------------------------------------------------------------------

\section{Implementation of the Hall numerical scheme}
\label{sec:apA}

The Hall-MHD scheme that we use in \NIR (with the HLLD solver) and in \FTD is based on the HDS of \citet{OSullivan2007} and follows the same procedure as described in appendix~A of \citet{Bai2014}.
The scheme is based on operator splitting the Hall electric field $\mathcal{E}_{\rm H} \equiv \etaH J \times \hat{b}$.
We begin by computing the $\mathcal{E}_{{\rm H}x}$ component at the position $(i, j+\frac{1}{2}, k+\frac{1}{2})$
\begin{equation}
  \mathcal{E}_{{\rm H}x} = \frac{\etaH}{|\B|} (J_y^*B_z^* - J_z^*B_y^*)\,,
\end{equation}
where the superscript  $*$ means that the components of the current and the magnetic field have to be interpolated to the $\mathcal{E}_{{\rm H}x}$ position.
We then update $B_y$ and $B_z$ for a full time step using a CT step, but only with the component $\mathcal{E}_{{\rm H}x}$, that is,
\begin{equation} \label{eq:CT1}
    B_{y,i,j+\frac{1}{2},k}^{n+1} = B_{y,i,j+\frac{1}{2},k}^{n} + \frac{\Delta t}{S_{XZ}} \left( \mathcal{E}_{{\rm H}x,i,j+\frac{1}{2},k+\frac{1}{2}} \Delta X_{i,j+\frac{1}{2},k+\frac{1}{2}} - \mathcal{E}_{{\rm H}x,i,j+\frac{1}{2},k-\frac{1}{2}}  \Delta X_{i,j+\frac{1}{2},k-\frac{1}{2}}\right)\,,
\end{equation}
\begin{equation}\label{eq:CT2}
    B_{z,i,j,k+\frac{1}{2}}^{n+1} = B_{z,i,j,k+\frac{1}{2}}^{n} - \frac{\Delta t}{S_{XY}} \left( \mathcal{E}_{{\rm H}x,i,j+\frac{1}{2},k+\frac{1}{2}} \Delta X_{i,j+\frac{1}{2},k+\frac{1}{2}} - \mathcal{E}_{{\rm H}x,i,j-\frac{1}{2},k+\frac{1}{2}}  \Delta X_{i,j-\frac{1}{2},k+\frac{1}{2}}\right)\,,
\end{equation}
where $S_{XZ}$ and $S_{XY}$ denote the area of the cell faces at which the magnetic fields $B_{y,i,j+\frac{1}{2},k}$ and $B_{z,i,j,k+\frac{1}{2}}$ are defined at, respectively.
With the updated values $B_y^{n+1}$ and $B_z^{n+1}$, we compute $\mathcal{E}_{{\rm H}y}$ and do the update of $B_x^{n+1}$ and $B_z^{n+1}$ using the equivalent of equations \ref{eq:CT1} and \ref{eq:CT2}.
In the same manner we compute $\mathcal{E}_{{\rm H}z}$ with the new updated magnetic field components.
In \FTD, with the computed $\mathcal{E}_{\rm H}$ we do an update of $B^{n} \rightarrow B^{n+1}$ using the sum of all the electric fields., that is,
\begin{equation}
    \mathcal{E}_{\rm T} = \mathcal{E}_{\rm I} + \mathcal{E}_{\rm H} + \mathcal{E}_{\rm O} + \mathcal{E}_{\rm A}\,,
\end{equation}
while in \NIR, the update is simply operator-split, which means that the state of the magnetic field before the Hall-specific update does not have to be stored. The described update naturally lends itself to sub-stepping (see \Sec{subcyc}) in both codes.

The \citeauthor{2008JCoPh.227.6967T} scheme that is used with HLL fluxes in \NIR closely follows the implementation by \citet{Lesur2014} in the \textsc{pluto} code. We found that, as an empirical requirement for stability when interpolating the face-centered electric fields to the cell edges, the (more accurate) up-winded \citet{Gardiner2008} interpolation has to be sacrificed in favor of plain arithmetic averages (also G.~Lesur, priv. comm.). While arithmetic averaging is unstable in the context of the more accurate HLLD \citep{Flock2010}, it is tolerable with the intrinsically more diffusive HLL solver.

% -------------------------------------------------------------------------------

\section{Testing the Hall module}
\label{sec:apB}

\subsection{Shock test} \label{subsec:shock} % ---

\begin{figure}\label{fig:shock}
  \centering\includegraphics[width=0.9\textwidth]{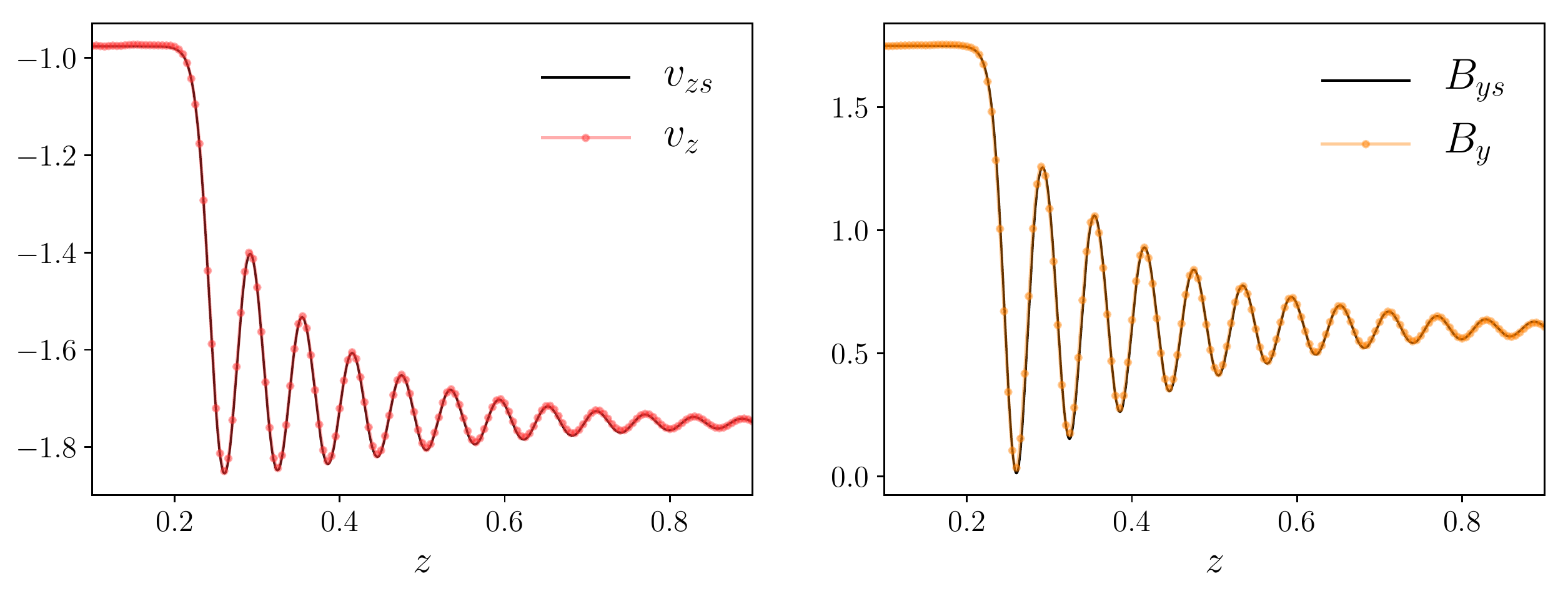}
  \caption{Solutions of the shock test problem of the fluid velocity (left panel) and magnetic field (right panel). Black solid lines show the analytic steady-state shock solution, while data points are obtained from the numerical solution at $t=2.7$, when the system has relaxed.}
\end{figure}

As a baseline non-linear test, we perform the shock tube test problem described in \citet{OSullivan2006}.
The shock propagates under the combined effect of ambipolar and Ohmic diffusion, along with the Hall effect.
Because the original problem is formulated for three-fluid MHD, in order to obtain the correct diffusivity coefficients, we have to solve a multifluid system of equations \citep[see][]{OSullivan2006}.
Using the multifluid version of \FTD, we update the densities of the two charged species via a continuity equation and we solve their velocities, $\vi$, assuming steady-state in the momentum equation. This yields
\begin{equation}
  \mathcal{E} + (\vi - \vg) \times \B
              - \frac{\rho_{\rm g} \K_i}{\alpha_i}(\vi - \vg) = 0\,,
\end{equation}
where $\vg$ denotes the velocity of the neutrals, and $\alpha_i$ is the charge-to-mass ratio of the species.
Assuming constant coefficients, $\K_i$, the velocity of each of the charged species can be obtained via
\begin{equation}
  (\vi - \vg )\, \left( |\B|^2+\xi^2 \right) =
  \displaystyle{\frac{(\mathcal{E} \cdot \B)}{\xi}}\B
                + \xi\,\mathcal{E} + \mathcal{E} \times \B\,,
\end{equation}
where $\xi = \rho_{\rm g} \K_i/\alpha_i$ and $\mathcal{E}$ was defined in  \Equation{EMF} in \Sec{equations}.
In our case, the diffusion coefficients are $\etaA \equiv \RA$, $\etaH \equiv \RH$ and $\etaO \equiv \RO$, where $\RA$, $\RH$ and $\RO$ are computed using equations (10) and (11) in \citet{OSullivan2006}.
The initial conditions are such that the left and right states are given by
\begin{eqnarray}
  \rho_{\rm L} = 1.7942\,,\;
  v_{z \rm L} = -0.9759\,,\;
  v_{y \rm L} = -0.6561\,,\; &&
  B_{y \rm L} = 1.74885\,,\;
  \rho_{1 \rm L} =  8.9712\ee{-8}\,,\;
  \rho_{2 \rm L} = 1.7942 \ee{-3}\,, \quad\\
  \rho_{\rm R} = 1.0\,,
  v_{z\rm R} = -1.751\,,
  v_{y\rm R} = 0.0\,, &&
  B_{y\rm R} = 0.6\,,
  \rho_{1\rm R} = 5\ee{-8}\,
  \rho_{2\rm R} = 1\ee{-3}\,,
\end{eqnarray}
respectively. The sound speed is $\cs = 0.1$ and $B_z = 1.0$ for both states, while $v_x = B_x = 0.0$. The coefficients for the charged species are
\begin{equation}
  \alpha_1 = -2\ee{9}\,,\;
  \alpha_2 = 1\ee{5}\,,\;
  \K_1 = 4\ee{2}\,,\;
  \K_2 = 2.5\ee{6}\,.
\end{equation}
The shock propagates along the $z$~direction in a domain of unit size, using 500 and 1000 cells, respectively. The initial discontinuity is located at $z=0.25$, and the boundary conditions are fixed at the initial state.
In \Fig{shock}, we show the numerical solution at $t=2.7$.
The L1~error for a grid resolution of $\Delta z = 2\ee{-2}$ is $e_1= 3.66 \ee{-2}$ and with a resolution of $\Delta z = 1\ee{-3}$, it is $e_1 = 9.8 \ee{-3}$, giving the scaling $e_1 \sim \Delta z^{1.9}$ which is close to the expected second-order convergence.

\subsection{Linear Wave Convergence} \label{subsec:obl} % ---

\begin{figure}
  \centering
  \includegraphics[width=0.31\textwidth]{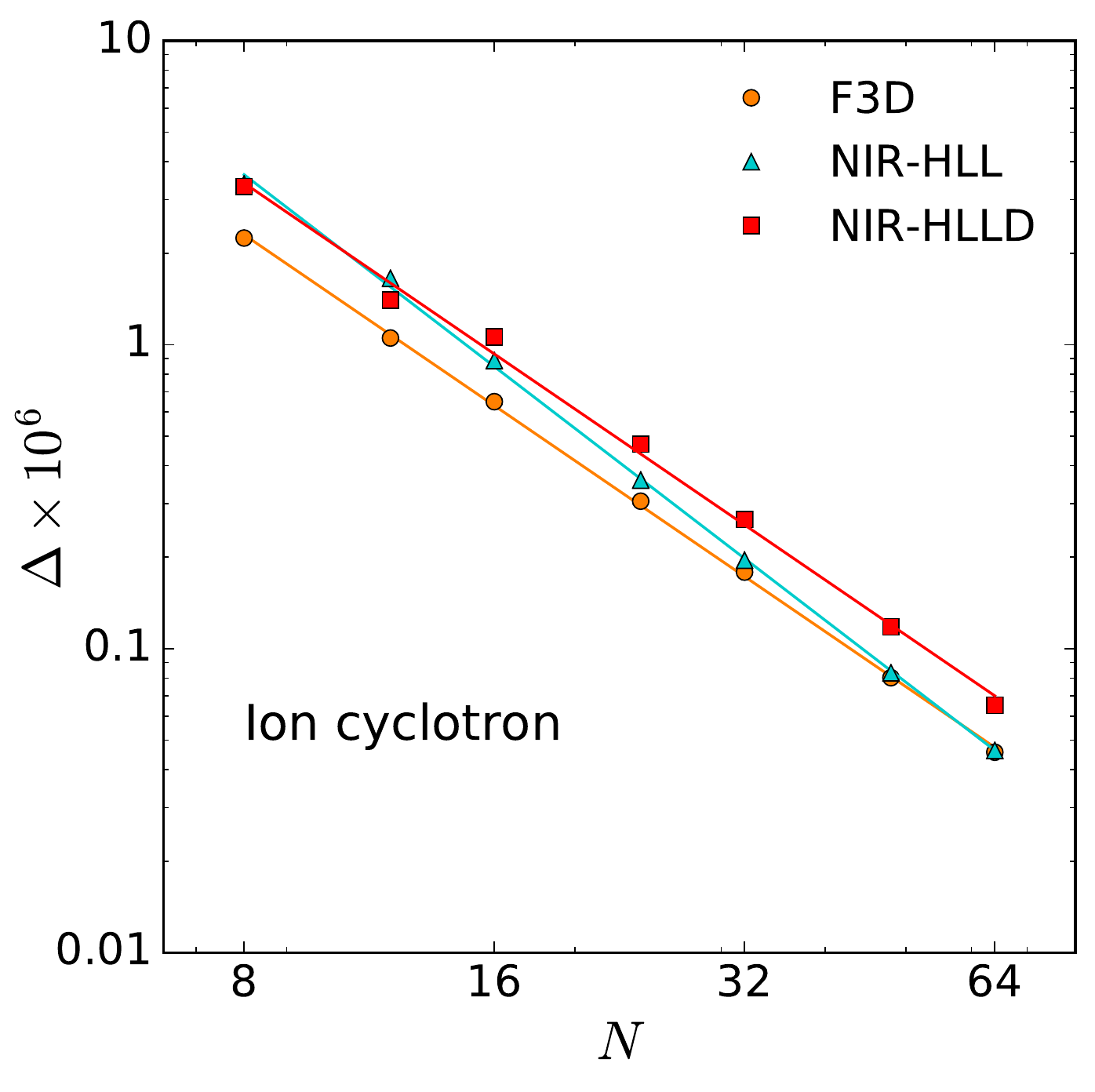}
  \includegraphics[width=0.31\textwidth]{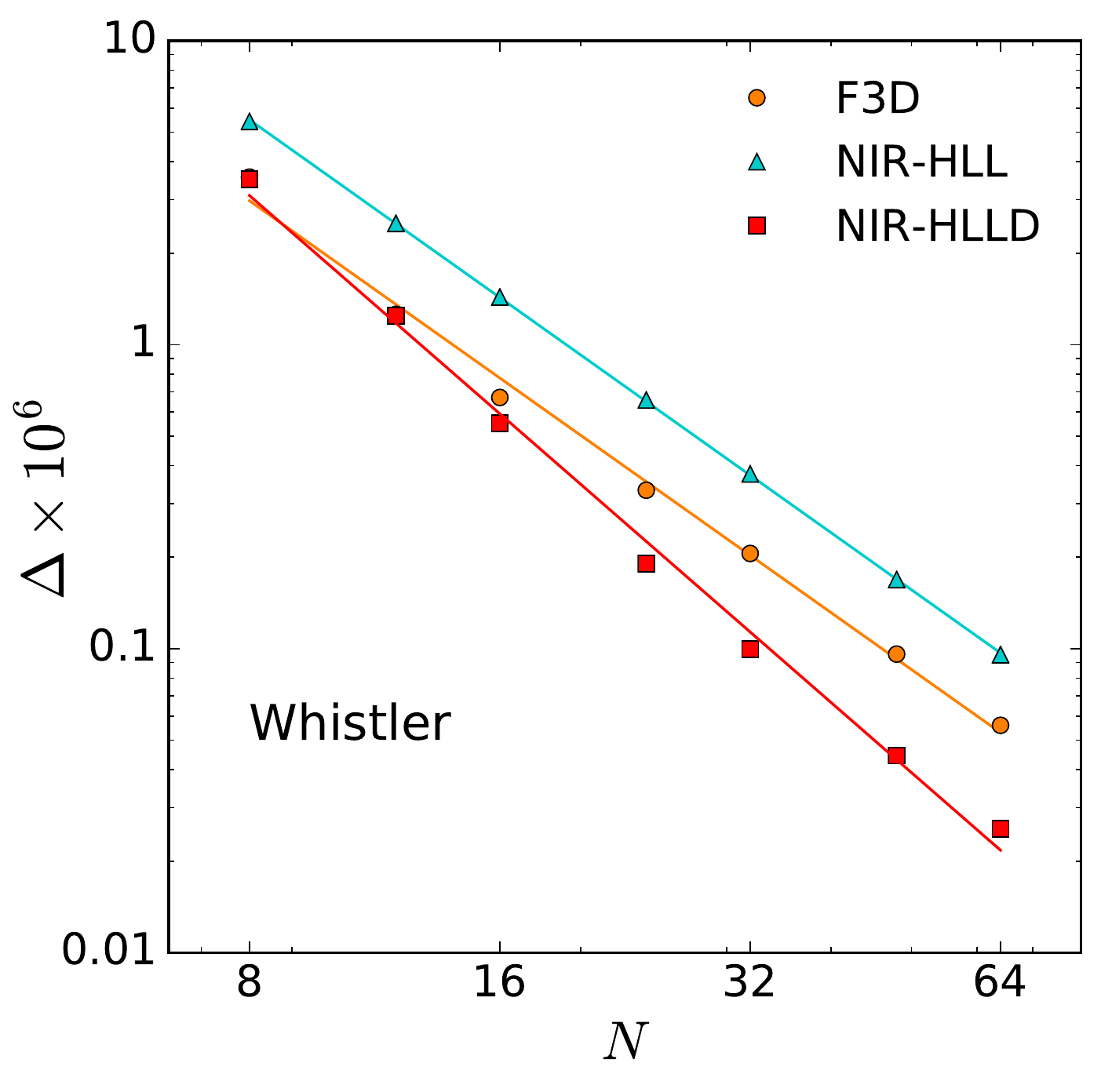}
  \includegraphics[width=0.31\textwidth]{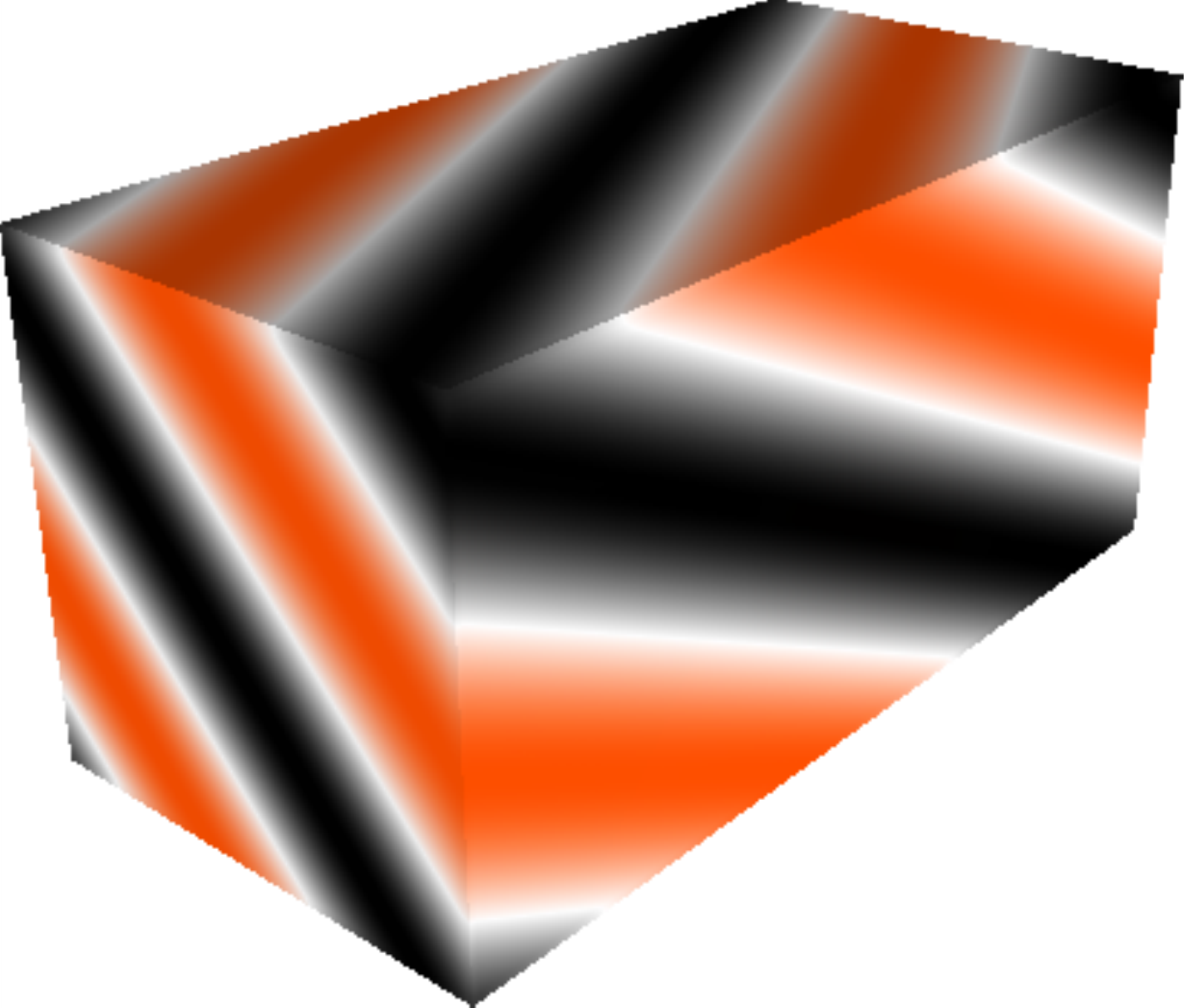}
  \caption{Linear wave convergence test for an oblique circular-polarized Alfv{\'e}n wave. \emph{Left and center panel:} Truncation error, where solid lines mark least square fits to the function $\Delta$, defined via $\Delta^2 \equiv \sum_s (\delta q_s)^2$, where $\delta q_s$ is the L-1 error of the s-component of the magnetic field. \emph{Right panel:} illustration of the oblique wave mode.}\label{fig:obl}
\end{figure}

In this section, we consider the linearized Hall-MHD equations of an incompressible fluid with only the Lorentz Force in the momentum equation.
Let us assume a background magnetic field $\B_0 = B_0 \hat{x}_1$ subject to the fluctuations $\delta \B = (0,\delta B_2, \delta B_3)$ and a perturbed velocity field  $\delta \V = (0, \delta v_2, \delta v_3)$. All perturbations have the form $\delta f = \delta f_0 {\rm exp}[i\, (\omega t - {\bf{k}} \cdot {\bf{x}})]$ along the direction $x_1$, that is, ${\bf{k}} = \pm k\hat{x}_1$.
It can be shown that
\begin{equation} \label{eq:eigen_amp}
  \delta v_{2,3} = - \frac{{\bf{k}} \cdot \B_0}{\mu_0 \rho_0 \omega}
                 \,\delta B_{2,3}\,,
\end{equation}
where $\delta B_{2,3}$ is the non-trivial solution of the system $A\,\delta B = 0$, with $A$ being the matrix
\begin{eqnarray}
 A & = & \left[
\begin{array}{cc}
  i\,(-\omega^2 + v_A^2k^2)  &  -\omega\,\etaH k^2        \\
  \omega\,\etaH k^2          &  i\,(-\omega^2 + v_A^2k^2) \\
\end{array}\right]\,,
\end{eqnarray}
and where  $v_A = B_0/\sqrt{\mu_0 \rho_0}$ is the Alfv\'en speed, and $\etaH$ is considered constant and normalized by $B_0$.
The system has four independent solutions, which correspond to two circular polarized waves propagating along $\pm {x}_1$.
For non-zero $\etaH$, these waves are commonly know as the whistler mode and the ion-cyclotron mode, respectively.
Characteristically, these are dispersive waves, that is, the phase velocity, $\omega/k$, depends on the wavenumber. More precisely,
\begin{equation}
  \frac{\omega}{k} = \frac{\etaH}{2} ({\bf{k}} \cdot \bb ) \pm \bigg[ \frac{1}{4}(k\etaH)^2 + v_A^2 \bigg]^{\frac{1}{2}}\,,
\end{equation}
with $\bb$ the unit vector in the direction of $\B_0$.

In order test the newly developed Hall-MHD modules, we perform a convergence study of an oblique wave propagation following a similar ansatz as the one presented in \citet{Gardiner2008} for circularly polarized waves in the ideal-MHD regime.
We define a Cartesian periodic box with dimensions $(3,1.5,1.5)$, resolved by  $(2N,N,N)$ grid cells in $x$, $y$, and $z$, respectively.
The propagation is along the oblique coordinate $x1 = k_xx + k_yy + k_z z$, with
\begin{equation}
  k_x = 2\pi \cos(\alpha)\cos(\beta)\,,\quad
  k_y = 2\pi \cos(\alpha)\sin(\beta)\,,\quad \text{and} \quad
  k_z = 2\pi \cos(\alpha)\,.
\end{equation}
We moreover set $\alpha=2/3$ and $\beta=2/\sqrt{5}$, in order to fit one wavelength $\lambda = 1$ inside the box.
The initial velocity field is $\vec{v} = (0,10^{-6}\cos(2\pi x_1),10^{-6}\sin(2\pi x_1))$ and the initial magnetic field is computed via a vector potential in such a way that $\B$ satisfies \Eqn{eigen_amp} with a background field $\B_0 = 1.0\hat{x}_1$.
The gas is treated as isothermal where $P_0 = \cs^2\rho_0$, with an initial density $\rho_0 = 1.0$ and gas pressure $P_0=1.0$.
The Hall diffusion is set to $\etaH = 0.1$.

We compute the L1-error for the centered components of the magnetic field as defined in Equation~(70) by \citet{Gardiner2008}, and we plot the results for different resolutions in \Fig{obl}.
For the Whistler mode \NIR returns an error $e_1 \sim \Delta_z^{2.3}$ and $e_1 \sim \Delta_z^{1.94}$, with the HLLD and HLL solvers, respectively.
This scaling is different for the Ion cyclotron mode, where $e_1 \sim \Delta_z^{1.87}$ with HLLD solver and $e_1 \sim \Delta_z^{2.1}$ with the HLL scheme.
In the case of \FTD, the error goes as $e_1 \sim\Delta z^{1.94}$ for the Whistler mode and $e_1 \sim\Delta z^{1.87}$ for the ion cyclotron mode.
All the results are reasonably close to the expected second order convergence of the implemented schemes.

\begin{figure}
        \centering
        \includegraphics[width=0.4\textwidth]{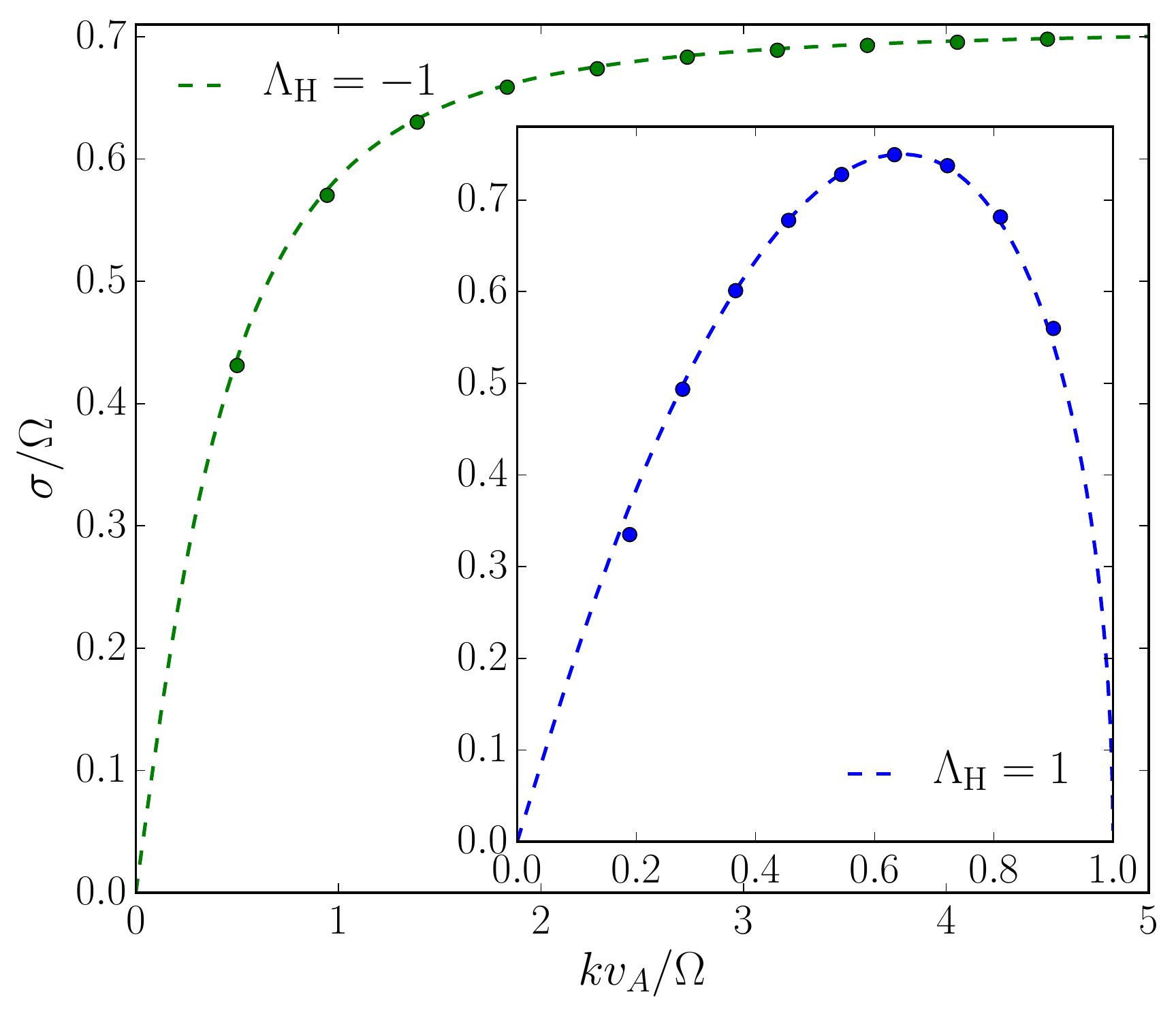}\hspace{1cm}
        \includegraphics[width=0.4\textwidth]{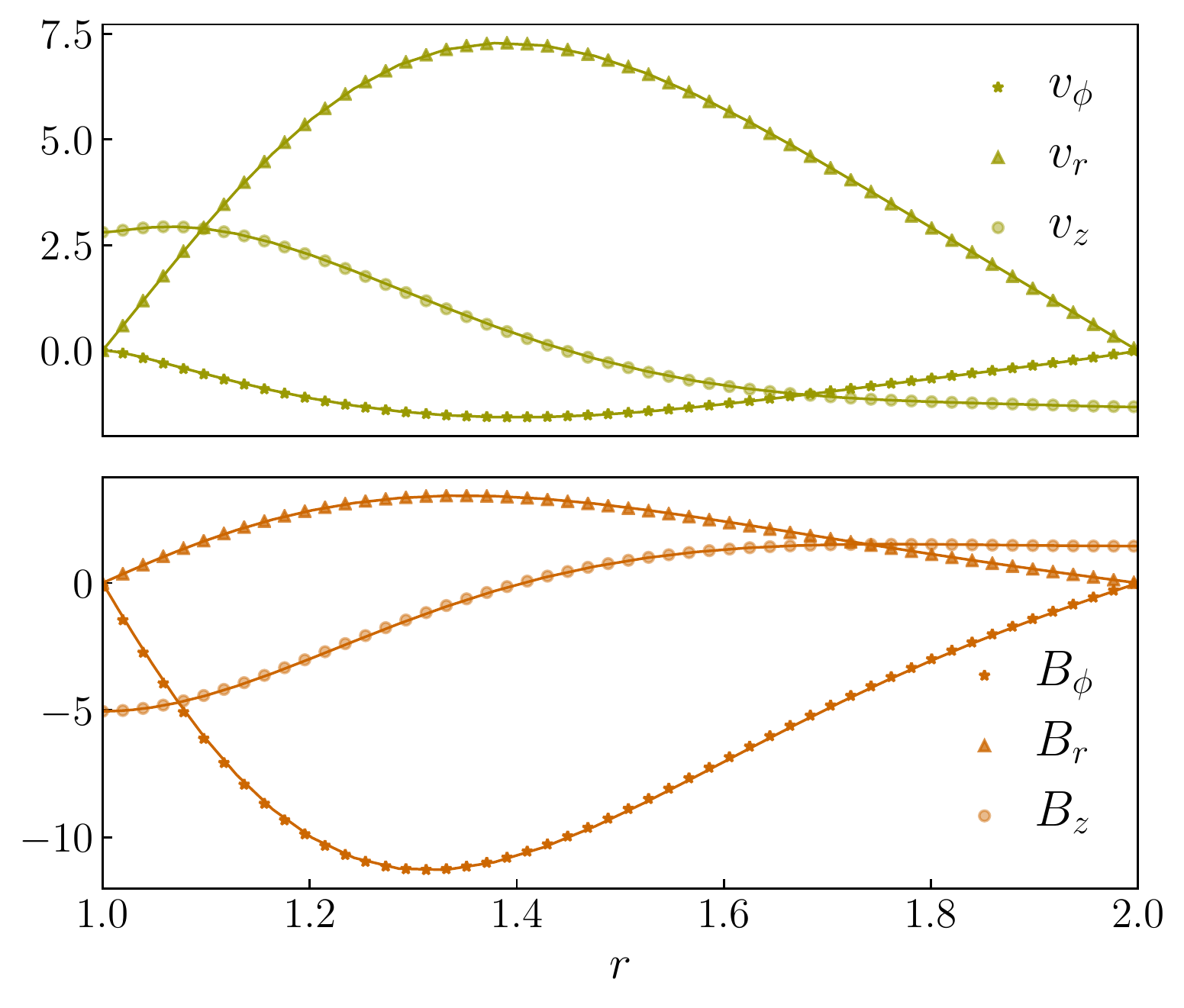}
        \caption{\emph{Left panel:} Analytic (dashed lines) and numerical (points) growth rates of the linear, local MRI. \emph{Right panel:} Analytic (lines) and numerical (symbols) eigenvectors of linear, global cylindrical MRI modes. The velocities are normalized by a factor $10^{-6}$ and the magnetic field components by a factor $10^{-5}$. Both results were obtained with \FTD.} \label{fig:lineargrowth}
\end{figure}

\subsection{Linear MRI growth --- local modes} \label{subsec:lineargrowth} % ---

We now study the growth rate of linear MRI modes under the Hall effect.
We use an axisymmetric, two-dimensional cylindrical domain with a grid of $64 \times 64$ cells, where the radial domain is fixed to $r\in [0.8,1.2]$.
We adopt a strategy where we initialize simulations such that the vertical box size matches the wavelength of a given individual mode.
In this way, we run one simulation for each mode in a box with $L_z = \lambda \equiv 2\pi/k_z$, maintaining the same effective resolution for all the modes.

We use a similar initial condition as \citet{Sano2002}, where the plasma-$\beta$ parameter is set to $\beta = 800$, $\rho_0 = 1.0$, $\cs = 0.1$, and we apply a Keplerian background velocity field.
We then add perturbations to the azimuthal and radial velocities of the form $v_0 \cos(kz)$ and $v_0 \sin(kz)$, respectively, where $v_0=10^{-6}\cs$.
The Hall diffusivity coefficient, $\etaH$, is defined in such a way that the Elsasser numbers are $\Lambda_{\rm H} = 1$ when $\B \cdt \Om > 0$ and $\Lambda_{\rm H} = -1$ when $\B \cdt \Om < 0$.
We integrate for two orbits at $r=1$, and we measure the growth rate between  $t=1.5-1.9$ orbits at the same radius.
By means of an exponential fit to the maximum value of the radial magnetic field at $r=1$, we obtain the numerical solutions shown in the left panel of \Fig{lineargrowth}, which agrees with the analytic solution to within 3\% for the critical mode and 0.02\% for the maximum growth rate.
We attribute deviations to the discrepancy between the local Cartesian approximation and the cylindrical computational setup.

\subsection{Linear MRI growth --- global modes} \label{subsec:eigen} % ---

The eigenvectors that we show are the solutions of the linear Hall-MRI equation in a cylindrical coordinate system assuming axisymmetric perturbations.
The perturbed MHD system of equations is linear but not fully algebraic.
Thus, to compute the semi-analytic solutions we use a spectral code that expands the radial velocity and  magnetic field using Chebyshev polynomials.
The numerical setup is initialized using random white noise in the velocity field of the order $10^{-8}\cs$ in a global isothermal disk.
We adopt a box of size $(L_r,L_z) = [1,1]$ with $256 \tms 256$ cells, with periodic boundaries in the vertical and azimuthal directions.
The radial boundary conditions we apply are $ v_r = b_r = b_{\phi} = \partial_r b_z = \partial_r v_z = 0 $.
The disk is assumed to rotate with a Keplerian profile, and the initial conditions are
\begin{equation}
  \beta = 31250\,,\quad
  \rho = 1.0\,,\quad
  \cs = 0.25\,,\quad
  \etaO = 0.003\,, \quad \text{and} \quad
  \lH = 1.0\,,
\end{equation}
In the right panel of \Fig{lineargrowth}, we plot the eigenvectors obtained after 37 orbits.
We compute as well the growth rate of the modes via a linear fit of $\log (B_r(t))$.
We obtained a growth rate $\gamma = 0.0989$ with \FTDs, $\gamma = 0.0989$ ($\gamma = 0.0988$) with \NIRs HLLD (HLL). All the values are in excellent agreement with the expected analytic value of $\gamma_0 = 0.0989$ \citep[also cf.][]{Bethune2016a}.

% ------------------------------------------------------------------------------

%\bibliographystyle{aasjournal-hyperref}

\end{document}